\documentclass[sigconf]{acmart}

\usepackage[utf8]{inputenc}
\usepackage{booktabs,multirow}
\usepackage{enumitem}
\usepackage{color}
\usepackage{rotating}
\usepackage{amsmath}
\usepackage{array}
\usepackage{enumitem}
\usepackage[linesnumbered,algoruled,boxed,lined]{algorithm2e}
\usepackage{balance}
\usepackage[short]{optidef}

\newcommand{\ie}{{\it i.e.}}
\newcommand{\eg}{{\it e.g.}}

\newcommand{\ul}{\underline}{}

\newtheorem{theorem}{Theorem}[section]
\newcommand{\red}[1]{\textbf{\textcolor{red}{#1}}}
\newcommand{\blue}[1]{\ul{\textcolor{blue}{#1}}}

\AtBeginDocument{%
  }

\copyrightyear{2023}
\acmYear{2023}
\setcopyright{acmlicensed}\acmConference[SIGIR '23]{Proceedings of the 46th International ACM SIGIR Conference on Research and Development in Information Retrieval}{July 23--27, 2023}{Taipei, Taiwan}
\acmBooktitle{Proceedings of the 46th International ACM SIGIR Conference on Research and Development in Information Retrieval (SIGIR '23), July 23--27, 2023, Taipei, Taiwan}
\acmPrice{15.00}
\acmDOI{10.1145/3539618.3591704}
\acmISBN{978-1-4503-9408-6/23/07}

\begin{document}

\title{It's Enough: Relaxing Diagonal Constraints in Linear Autoencoders for Recommendation}
\author{Jaewan Moon}
\affiliation{
  \institution{Sungkyunkwan University}
  \country{Republic of Korea}}
\email{jaewan7599@skku.edu}

\author{Hye-young Kim}
\affiliation{
  \institution{Sungkyunkwan University}
  \country{Republic of Korea}}
\email{khyaa3966@skku.edu}

\author{Jongwuk Lee}\authornote{Corresponding author}
\affiliation{
  \institution{Sungkyunkwan University}
  \country{Republic of Korea}}
\email{jongwuklee@skku.edu}

\renewcommand{\shortauthors}{Jaewan Moon, Hye-young Kim, \& Jongwuk Lee}

\begin{abstract}
Linear autoencoder models learn an item-to-item weight matrix via convex optimization with \emph{L2 regularization} and \emph{zero-diagonal constraints}. Despite their simplicity, they have shown remarkable performance compared to sophisticated non-linear models. This paper aims to theoretically understand the properties of two terms in linear autoencoders. Through the lens of singular value decomposition (SVD) and principal component analysis (PCA), it is revealed that L2 regularization enhances the impact of high-ranked PCs. Meanwhile, zero-diagonal constraints reduce the impact of low-ranked PCs, leading to performance degradation for unpopular items. Inspired by this analysis, we propose simple-yet-effective linear autoencoder models using diagonal inequality constraints, called \emph{Relaxed Linear AutoEncoder} (RLAE) and \emph{Relaxed Denoising Linear AutoEncoder} (RDLAE). We prove that they generalize linear autoencoders by adjusting the degree of diagonal constraints. Experimental results demonstrate that our models are comparable or superior to state-of-the-art linear and non-linear models on six benchmark datasets; they significantly improve the accuracy of long-tail items. These results also support our theoretical insights on regularization and diagonal constraints in linear autoencoders.
\end{abstract}




\maketitle

\section{Introduction}\label{sec:introduction}
Over the last three decades, the field of recommender systems~\cite{JannachPRZ21} has been dedicated to helping users overcome information overload in various applications, \eg, Amazon, Netflix, and Bing News. \emph{Collaborative filtering (CF)}~\cite{GoldbergNOT92CF, HerlockerKBR99CFUserKNN} is a prevalent solution for building recommender systems due to its ability to uncover hidden collaborative signals from user-item interactions. Existing CF models can be broadly categorized into \emph{linear} and \emph{non-linear} approaches based on how they capture correlations among users/items. Linear models represent the user/item relationships through a linear combination. With the advent of deep learning, non-linear CF models utilize various neural networks, \ie, autoencoders~\cite{WuDZE16CDAE, LiangKHJ18MultVAE, ShenbinATMN20RecVAE, LobelLGC20RaCT} (AE), recurrent neural networks (RNN)~\cite{HidasiKBT15GRU4Rec, LiRCRLM17NARM}, transformers~\cite{KangM18SASRec, SunLWPLOJ19BERT4Rec}, and graph neural networks~\cite{Wang0WFC19NGCF, ChenWHZW20LRGCCF, 0001DWLZ020LightGCN, 0002JP21LTOCF, ShenWZSZLL21GFCF, KongKJ0LPK22HMLET} (GNN). They claim that non-linear models surpass linear models in capturing intricate and scarce collaborative signals, leading to better performance in various recommendation scenarios.

However, there are exciting research debates on evaluating linear and non-linear models. Recent studies~\cite{DacremaCJ19, SunY00Q0G20, abs-1905-01395, DacremaBCJ21} claim that the hyperparameters of baselines need to be tuned carefully, or the choice of evaluation metrics heavily affects a fair comparison. Surprisingly, well-tuned linear models, such as neighborhood-based models~\cite{HerlockerKBR99CFUserKNN, SarwarKKR01CFItemKNN}, simple graph-based models~\cite{CooperLRS14P3alpha, PaudelCNB17RP3beta}, and linear matrix factorization models~\cite{ZhouWSP08ALSWR, Koren08SVD++} have achieved competitive or significant gains over non-linear models. Recent linear models using item neighborhoods, \eg, EASE$^R$~\cite{Steck19EASE} and EDLAE~\cite{Steck20edlae}, have shown state-of-the-art performance results on large-scale datasets, \eg, ML-20M, Netflix, and MSD. In this paper, we thus delve deeper into the linear models using item neighborhoods, also known as \emph{linear autoencoders}~\cite{NingK11SLIM, Steck19EASE, Steck20edlae, SteckDRJ20ADMMSLIM, JeunenBG20CEASE, SteckL21Higher, VancuraAKK22ELSA}.

Given a user-item interaction matrix $\mathbf{X} \in \{0,1\}^{m \times n}$, linear autoencoders (LAE) learn an \emph{item-to-item weight matrix} $\mathbf{B} \in \mathbb{R}^{n \times n}$ so that the matrix product $\mathbf{X} \cdot \mathbf{B}$ reconstructs the original matrix $\mathbf{X}$. It takes  $\mathbf{X}$ as input and output, and $\mathbf{B}$ represents a single hidden layer to serve both an encoder and a decoder. Existing studies~\cite{NingK11SLIM, Steck19EASE, Steck20edlae, SteckDRJ20ADMMSLIM, JeunenBG20CEASE, SteckL21Higher, VancuraAKK22ELSA} formulate a convex optimization problem with two key terms: \emph{regularization} and \emph{zero-diagonal constraints}. (i) Regularization is widely used to prevent the models from overfitting, \ie, $\hat{\mathbf{B}}=\mathbf{I}$. While SLIM~\cite{NingK11SLIM} utilizes both L1 and L2 regularization, EASE$^R$~\cite{Steck19EASE} employs only L2 regularization to derive a closed-form solution and shows better performance. However, \cite{SteckDRJ20ADMMSLIM} utilizes the alternating directions method of multipliers (ADMM) to optimize the objective function of SLIM, showing that L1 regularization only affects the sparsity of the solution and has little impact on recommendation results. Moreover, EDLAE~\cite{Steck20edlae} proposes linear autoencoder models with advanced regularization derived from random dropout. (ii) The zero-diagonal constraints in $\mathbf{B}$ are designed to identify precisely the correlation between items by preventing self-correlation~\cite{NingK11SLIM, Steck20edlae}. Since it seems natural that the diagonal entries in $\mathbf{B}$ are unnecessary weights, existing studies~\cite{NingK11SLIM, Steck19EASE, Steck20edlae, SteckDRJ20ADMMSLIM, JeunenBG20CEASE, SteckL21Higher, VancuraAKK22ELSA} employ the zero-diagonal constraints to eliminate them. However, how the regularization and the zero-diagonal constraints affect recommendation is unexplored.

In this paper, we ask two underlying questions about linear autoencoder models: (i) \emph{How does each term in linear autoencoders, \ie, regularization and zero-diagonal constraints, affect item popularity?} (ii) \emph{Do the zero-diagonal constraints always help improve model performance}? To answer these questions, we conduct a theoretical analysis of various linear autoencoder models. We rigorously derive the relationship among four linear autoencoder models, \ie, LAE~\cite{HoerlK00Ridge}, EASE$^R$~\cite{Steck19EASE}, DLAE~\cite{Steck20edlae}, and EDLAE~\cite{Steck20edlae}. It is revealed that the solutions of EASE$^R$ and EDLAE include those of LAE and DLAE, respectively. Besides, their solutions are decomposed into two terms for regularization and zero-diagonal constraints. We represent each term as an eigenvalue decomposition form through the lens of singular value decomposition (SVD). Interestingly, they share the same eigenvectors derived from the gram matrix $\mathbf{X}^{\top}\mathbf{X}$ but show different tendencies in the eigenvalues: while regularization heavily affects high-ranked PCs, the diagonal constraints have more impact on low-ranked PCs, penalizing weak collaborative signals.

To analyze how each term is related to item popularity, we further observe the eigenvectors of $\mathbf{X}^{\top}\mathbf{X}$ using principal component analysis (PCA). Since $\mathbf{X}^{\top}\mathbf{X}$ is approximately proportional to the covariance matrix, high-ranked PCs represent strong collaborative signals, and low-ranked PCs are related to weak collaborative signals. It indicates that the high-ranked PCs are highly co-related to popular items, and the low-ranked PCs are associated with less popular items. Based on this observation, we can respond to the questions; (i) as the regularization term increases, it is biased toward strong collaborative signals and tends to recommend popular items. (ii) Meanwhile, because the zero-diagonal constraints mostly penalize the impact of the low-ranked PCs, it weakens the collaborative signals for less popular items. As a result, we conclude that relaxing the diagonal constraints helps mitigate popularity bias and recommend less popular items.

To this end, we propose novel linear autoencoder models called \emph{Relaxed Linear AutoEncoder} (RLAE) and \emph{Relaxed Denoising Linear AutoEncoder} (RDLAE). First, we formulate a new convex optimization problem using the diagonal inequality constraints. Surprisingly, it is derived that the solution form of RLAE is similar to that of EASE$^{R}$~\cite{Steck19EASE}, and it is possible to control the degree of the diagonal constraints. Inspired by DLAE~\cite{Steck20edlae}, we extend RLAE to RDLAE by employing dropout regularization and obtain a solution similar to EDLAE~\cite{Steck20edlae}. We then prove that our models generalize existing linear autoencoder models by controlling the hyperparameter for inequality constraints. Extensive experimental results demonstrate that our models are comparable to or even better than existing linear and non-linear models on six benchmark datasets with various evaluation protocols; they significantly improve the accuracy of long-tail items. These results also support our theoretical insights on regularization and diagonal constraints in linear autoencoders.

We summarize the main contributions of this work as follows.
\begin{itemize}[leftmargin=5mm]
     \item (Section~\ref{sec:analysis}) We analyze the solutions of existing linear autoencoder models, decomposed into two terms for regularization and zero-diagonal constraints. We conduct theoretical analyses through the lens of singular value decomposition (SVD) and principal component analysis (PCA) and then understand the effect of each term and the relationship for item popularity. In brief, the diagonal constraints can suppress the collaborative signals of unpopular items.

\vspace{0.5mm}
    \item (Section~\ref{sec:model}) We propose novel linear autoencoder models called \emph{Realxed Linear AutoEncoder} (RLAE) and \emph{Relaxed Denoising Linear AutoEncoder} (RDLAE). They introduce diagonal inequality constraints to mitigate the adverse effects of zero-diagonal constraints. We also prove that our models generalize existing linear autoencoder models. 

\vspace{0.5mm}
    \item (Section~\ref{sec:results}) Experimental results extensively demonstrate that our models perform competitively or better than state-of-the-art linear and non-linear models on six benchmark datasets. These results support our theoretical insights on regularization and diagonal constraints. Notably, our models achieve significant performance gains in long-tail items.
\end{itemize}

\section{Background}\label{sec:background}
\textbf{Notations}. Let a training dataset consist of $m$ users and $n$ items. In this paper, we assume implicit user feedback because it is more commonly used in various Web applications than explicit user feedback. Under this assumption, the user-item interaction matrix $\mathbf{X}$ is represented by a binary matrix, \ie, $\mathbf{X} \in \{0,1\}^{m \times n}$. If user $u$ has interacted with item $i$, then $x_{ui}=1$. $x_{ui}=0$ indicates no observed interaction between user $u$ and item $i$.

The goal of top-$N$ recommender models is to retrieve the top-$N$ items that the user is most likely to prefer. Existing linear models can be categorized into two directions: (1) the regression-based approach~\cite{NingK11SLIM, Steck19EASE, Steck20edlae, JeunenBG20CEASE, SteckL21Higher, VancuraAKK22ELSA}, also known as the linear autoencoder approach, and (2) the matrix factorization approach~\cite{HuKV08WMF1, PanZCLLSY08WMF2, ZhouWSP08ALSWR, Koren08SVD++}. Inspired by item-based neighborhood models, the linear autoencoder learns an item-to-item similarity matrix using the relationship between item neighborhoods. In contrast, the matrix factorization approach learns low-rank user and item matrices by factorizing the user-item matrix.

In this paper, we focus on addressing linear autoencoder models. They deal with the same matrix for input and output and learn the item-to-item weight matrix $\mathbf{B} \in \mathbb{R}^{n \times n}$. For inference, the linear autoencoder models then calculate the prediction score $s_{ui}$ using the inner product of two vectors.
\begin{equation}\label{eq:prediction}
    s_{ui} = \mathbf{X}_{u*} \cdot \mathbf{B}_{*i},
\end{equation}
where $\mathbf{X}_{u*}$ and $\mathbf{B}_{*i}$ refer to the row vector for user $u$ in $\mathbf{X}$ and the column vector for item $i$ in $\mathbf{B}$, respectively. Although it is possible to factorize the item-to-item weight matrix into two low-rank matrices~\cite{Steck20edlae, VancuraAKK22ELSA}, we mainly consider the full-rank matrix for $\mathbf{B}$.

\vspace{1mm}
\noindent
\textbf{Linear autoencoder (LAE)}. As the simplest model, the objective function of LAE is formulated with L2 regularization, equal to ridge regression~\cite{HoerlK00Ridge}.
\begin{equation}\label{eq:LAE_objective}
  \min_{\mathbf{B}} \|\mathbf{X} - \mathbf{X} \mathbf{B}\|_F^2 + \lambda \|\mathbf{B}\|_F^2,
\end{equation}

For LAE, we can easily derive the closed-form solution.
\begin{equation}\label{eq:LAE_solution}
    \begin{split}
        \hat{\mathbf{B}}_{LAE}   & = \left(\mathbf{X}^\top \mathbf{X} + \lambda \mathbf{I}\right)^{-1}\left(\mathbf{X}^\top \mathbf{X}\right) \\
                        & = \left(\mathbf{X}^\top \mathbf{X} + \lambda \mathbf{I}\right)^{-1}\left(\mathbf{X}^\top \mathbf{X} + \lambda \mathbf{I} - \lambda \mathbf{I}\right) \\
                        & = \mathbf{I} - \left(\mathbf{X}^\top \mathbf{X} + \lambda \mathbf{I}\right)^{-1} \lambda \mathbf{I} \\
    \end{split}
\end{equation}

Here, $\mathbf{X}^{\top}\mathbf{X} \in \mathbb{R}^{n \times n}$ represents a gram matrix approximately proportional to the covariance matrix, and $\mathbf{I}$ is an identity matrix. When $\lambda=0$, it comes to a trivial solution, \ie, $\hat{\mathbf{B}}_{LAE} = \mathbf{I}$. Therefore, it is natural to choose a positive value of $\lambda$. Although the L2 regularization prevents a trivial solution, it does not perfectly decouple the self-correlation in $\mathbf{B}$.

\vspace{1mm}
\noindent
\textbf{EASE$^{R}$}~\cite{Steck19EASE}. It introduces diagonal constraints on $\mathbf{B}$, allowing us to avoid a trivial solution regardless of the L2 regularization. The convex optimization problem of EASE$^{R}$ employs two components: (1) an objective function with L2 regularization and (2) zero-diagonal constraints.
\begin{equation}\label{eq:ease_objective}
    \min_{\mathbf{B}} \|\mathbf{X} - \mathbf{X} \mathbf{B}\|_F^2 + \lambda \|\mathbf{B}\|_F^2 \ \ s.t. \ \ \text{diag}(\mathbf{B})=0
\end{equation}
where $\text{diag}(\mathbf{B})$ is the vector of diagonal entries for the matrix $\mathbf{B}$. If the zero-diagonal constraints do not exist, it is equal to Eq.~\eqref{eq:LAE_solution}.

The optimization problem is transformed by forming Lagrangian multipliers to account for the zero-diagonal constraints.
\begin{equation}\label{eq:ease_lagrangian}
    \min_{\mathbf{B}} \|\mathbf{X} - \mathbf{X} \mathbf{B}\|_F^2 + \lambda \|\mathbf{B}\|_F^2 + \boldsymbol{\mu}^{\top} \cdot \text{diag}(\mathbf{B})
\end{equation}
where $\boldsymbol{\mu}=(\mu_{1}, \ldots, \mu_{n})^{\top}$ denotes the vector of Lagrangian multipliers. Although it requires additional parameters, it is possible to derive the closed-form solution by minimizing Eq.~\eqref{eq:ease_lagrangian}.
\begin{equation}\label{eq:ease_solution}
    \begin{split}
    \hat{\mathbf{B}}_{EASE^{R}} &= \left(\mathbf{X}^{\top} \mathbf{X} + \lambda \mathbf{I}\right)^{-1}\left( \mathbf{X}^{\top}\mathbf{X} - \text{diagMat}\left(\boldsymbol{\mu}\right) \right) \\
                                &= \mathbf{I} - \mathbf{P} \cdot  \text{diagMat}\left( \textbf{1} \oslash \text{diag}(\mathbf{P}) \right) \\
    \end{split}
\end{equation}
where $\mathbf{P} = (\mathbf{X}^{\top} \mathbf{X} + \lambda \mathbf{I})^{-1}$, $\textbf{1}$ and $\oslash$ are a vector of ones and the element-wise division operator, respectively. Here, the Lagrangian multipliers are determined by satisfying the equality constraints, \ie, $\text{diag}(\mathbf{B})=0$.

The solution of EASE$^{R}$ can be divided into two terms: \emph{regularization} and \emph{diagonal constraints}. The former is \emph{equivalent} to the solution of LAE, and the latter represents zero-diagonal constraints.
\begin{equation}\label{eq:ease_solution3}
    \hat{\mathbf{B}}_{EASE^{R}}  = \mathbf{P} \cdot \left(\mathbf{X}^{\top} \mathbf{X}\right) - \mathbf{P} \cdot \text{diagMat}\left(\boldsymbol{\mu}\right)
\end{equation}

We observe that the zero-diagonal constraints are the product of two matrices $-\mathbf{P}$ and $\text{diagMat}\left(\boldsymbol{\mu}\right)$. In Section~\ref{sec:analysis}, we will further analyze the impact of diagonal constraints.

\vspace{1mm}
\noindent
\textbf{DLAE and EDLAE}~\cite{Steck20edlae}. A recent study~\cite{Steck20edlae} points out that the LAE tends to overfit the identity matrix because it is trained with the same features for input and output. To address this problem, \cite{Steck20edlae} utilizes random dropout denoising as an effective regularizer. It helps models predict one feature from the other in the input. As the number of training epochs with random dropout is close to infinite, the stochastic dropout has converged. Interestingly, it is asymptotically equivalent to L2 regularization. Given a dropout probability $p$, we formulate the objective function of the denoising linear autoencoder (DLAE).
\begin{equation}\label{eq:DLAE_objective}
    \min_{\mathbf{B}} \|\mathbf{X}-\mathbf{X} \mathbf{B}\|_F^2 + \| \boldsymbol{\Lambda}^{1 / 2} \cdot \mathbf{B}\|_F^2
\end{equation}
where $\Lambda = \frac{p}{1-p} \cdot \text{diagMat}\left(\text{diag}\left( \mathbf{X}^{\top} \mathbf{X}\right) \right) + \lambda$. The solution form of DLAE is equal to LAE except for regularization.
\begin{equation}\label{eq:DLAE_solution}
    \hat{\mathbf{B}}_{DLAE}   = \left(\mathbf{X}^\top \mathbf{X} + \boldsymbol{\Lambda}\right)^{-1}\left(\mathbf{X}^\top \mathbf{X}\right)  = \mathbf{I} - \left(\mathbf{X}^\top \mathbf{X} + \boldsymbol{\Lambda}\right)^{-1} \boldsymbol{\Lambda}
\end{equation}

Although DLAE utilizes dropout-based regularization, it does not entirely prevent a weight matrix from overfitting toward the identity matrix. To alleviate this problem, DLAE is extended to incorporate zero-diagonal constraints, called the emphasized denoising linear autoencoder (EDLAE). The convex optimization problem and the solution of EDLAE are as follows.
\begin{equation}\label{eq:edlae_objective}
    \min_{\mathbf{B}} \|\mathbf{X}-\mathbf{XB}\|_F^2 + \| \boldsymbol{\Lambda}^{1 / 2} \cdot \mathbf{B}\|_F^2 \ \ s.t. \ \ \text{diag}(\mathbf{B})=0
\end{equation}
\begin{equation}\label{eq:edlae_solution}
    \hat{\mathbf{B}}_{EDLAE} = \left(\mathbf{X}^{\top} \mathbf{X} + \boldsymbol{\Lambda}\right)^{-1}\left( \mathbf{X}^{\top}\mathbf{X} - \text{diagMat}\left(\boldsymbol{\mu}\right) \right)
\end{equation}

DLAE and EDLAE~\cite{Steck20edlae} improve LAE and EASE$^{R}$~\cite{Steck19EASE} by tuning for L2 regularization with random dropout. Notably, EASE$^{R}$ and EDLAE also further consider the diagonal constraints more effectively. However, it remains unanswered: How do L2 regularization and zero-diagonal constraints affect recommendation?

\section{Theoretical Analysis}\label{sec:analysis}
In this section, we investigate the effects of L2 regularization and diagonal constraints. Through singular value decomposition (SVD), the matrix $X$ is decomposed into three matrices.
\begin{equation}\label{eq:svd}
\mathbf{X}=\mathbf{U} \boldsymbol{\Sigma} \mathbf{V}^{\top}
\end{equation}
where $\mathbf{U}$ and $\mathbf{V}$ are unitary matrices, \ie, $\mathbf{U}^{\top} = \mathbf{U}^{-1}$, $\mathbf{V}^{\top} = \mathbf{V}^{-1}$. Also, $\boldsymbol{\Sigma}$ is the diagonal matrix for singular values. Assuming $m > n$, let $\text{diag}(\boldsymbol{\Sigma})$ denote the vector ($\sigma_{1}, \ldots ,\sigma_{n})$. The gram matrix $\mathbf{X}^{\top} \mathbf{X}$ can be rewritten as an eigenvalue decomposition form by replacing $\mathbf{X}$ with Eq.~\eqref{eq:svd}:
\begin{equation}\label{eq:gram_svd}
\mathbf{X}^{\top} \mathbf{X} = {\left( \mathbf{U} \boldsymbol{\Sigma} \mathbf{V}^{\top} \right)}^{\top} \left( \mathbf{U} \boldsymbol{\Sigma} \mathbf{V}^{\top} \right) = \mathbf{V} \left( \boldsymbol{\Sigma}^{\top} \boldsymbol{\Sigma} \right) \mathbf{V}^{\top} 
\end{equation}

We then analyze the closed-form solution of EASE$^{R}$. We decouple the L2 regularized objective and the zero-diagonal constraints in $\mathbf{B}$ as in Eq.~\eqref{eq:ease_solution3}. Note that the solution of EDLAE can also be decomposed into two terms, \ie, the solution of DLAE and the zero-diagonal constraints by replacing $\lambda \mathbf{I}$ with $\boldsymbol{\Lambda}$.

The closed-form solution of LAE in Eq.~\eqref{eq:LAE_solution}, \ie, the L2 regularized objective, can be rewritten as follows. Here, we utilize a trick using $\mathbf{VV}^{\top} = \mathbf{I}$.
\begin{equation}\label{eq:ridge_svd}
\begin{split}
\hat{\mathbf{B}}_{LAE} & = \left(\mathbf{X}^{\top} \mathbf{X}+\lambda \mathbf{I}\right)^{-1}\left(\mathbf{X}^{\top} \mathbf{X}\right) \\
                & = \left( \mathbf{V} \left(\boldsymbol{\Sigma}^{\top}\boldsymbol{\Sigma}\right) \mathbf{V}^{\top} + \lambda \mathbf{I} \right)^{-1} \left( \mathbf{V} \left(\boldsymbol{\Sigma}^{\top}\boldsymbol{\Sigma}\right) \mathbf{V}^{\top} \right) \\
                & = \left( \mathbf{V} \left(\boldsymbol{\Sigma}^{\top} \boldsymbol{\Sigma}\right) \mathbf{V}^{\top} + \mathbf{V} \mathbf{V}^{\top} \lambda \mathbf{I}\mathbf{V} \mathbf{V}^{\top} \right)^{-1} \left( \mathbf{V} \left(\boldsymbol{\Sigma}^{\top} \boldsymbol{\Sigma}\right) \mathbf{V}^{\top} \right) \\
                & = \left(\mathbf{V} \left( \left(\boldsymbol{\Sigma}^{\top} \boldsymbol{\Sigma}\right) + \mathbf{V}^{\top} \lambda \mathbf{I} \mathbf{V}\right) \mathbf{V}^{\top} \right)^{-1} \left( \mathbf{V} \left( \boldsymbol{\Sigma}^{\top} \boldsymbol{\Sigma}\right) \mathbf{V}^{\top} \right) \\
                & = \mathbf{V} \left( \left(\boldsymbol{\Sigma}^{\top} \boldsymbol{\Sigma}\right) + \mathbf{V}^{\top} \lambda \mathbf{I} \mathbf{V} \right)^{-1} \mathbf{V}^{\top} \left( \mathbf{V} \left(\boldsymbol{\Sigma}^{\top} \boldsymbol{\Sigma}\right) \mathbf{V}^{\top} \right) \\
                & = \mathbf{V} \left( \left(\boldsymbol{\Sigma}^{\top} \boldsymbol{\Sigma}\right) + \lambda \mathbf{I}\right)^{-1} \left(\boldsymbol{\Sigma}^{\top} \boldsymbol{\Sigma}\right) \mathbf{V}^{\top} \\
\end{split}
\end{equation}

The eigenvalue decomposition of $\hat{\mathbf{B}}_{LAE}$ is as follows. The diagonal matrix ${(\boldsymbol{\Sigma}^{\top} \boldsymbol{\Sigma} + \lambda \mathbf{I})}^{-1} (\boldsymbol{\Sigma}^{\top} \boldsymbol{\Sigma})$ represents the eigenvalues of $\hat{\mathbf{B}}_{LAE}$.
\begin{equation}
\label{eq:lae_svd}
\hat{\mathbf{B}}_{LAE} = \mathbf{V}\text{diag} \left(\frac{\sigma_{1}^{2}}{\sigma_{1}^{2}+\lambda}, \ldots ,\frac{\sigma_{n}^{2}}{\sigma_{n}^{2}+\lambda} \right)\mathbf{V}^{\top}
\end{equation}

Similarly, we derive the alternative form for the zero-diagonal constraints in EASE$^{R}$.
\begin{equation}\label{eq:diagonal_svd}
    \begin{split}
    \hat{\mathbf{B}}_{EASE^{R}} - \hat{\mathbf{B}}_{LAE}  & = -\left(\mathbf{X}^{\top} \mathbf{X} + \lambda \mathbf{I}\right)^{-1} \text{diagMat}\left(\boldsymbol{\mu}\right) \\
                                            % & = -\left( \mathbf{V} \left(\boldsymbol{\Sigma}^{\top}\boldsymbol{\Sigma}\right) \mathbf{V}^{\top} + \lambda \mathbf{I}\right)^{-1} \text{diagMat}\left(\boldsymbol{\mu}\right) \\
                                            % & =  -\left( \mathbf{V} \left(\boldsymbol{\Sigma}^{\top} \boldsymbol{\Sigma}\right) \mathbf{V}^{\top} + \mathbf{V} \mathbf{V}^{\top} \lambda \mathbf{I}\mathbf{V} \mathbf{V}^{\top} \right)^{-1} \text{diagMat}\left(\boldsymbol{\mu}\right) \\
                                            % & = -\left(\mathbf{V} \left( \left(\boldsymbol{\Sigma}^{\top} \boldsymbol{\Sigma}\right) + \mathbf{V}^{\top} \lambda \mathbf{I} \mathbf{V}\right) \mathbf{V}^{\top} \right)^{-1} \text{diagMat}\left(\boldsymbol{\mu}\right) \\
                                            & = -\mathbf{V} {\left( \left(\boldsymbol{\Sigma}^{\top} \boldsymbol{\Sigma}\right) + \lambda \mathbf{I} \right)}^{-1} \mathbf{V}^{\top} \text{diagMat}\left(\boldsymbol{\mu}\right) \\
    \end{split}
\end{equation}

We also represent the eigenvalue decomposition for $\hat{\mathbf{B}}_{EASE^{R}} - \hat{\mathbf{B}}_{LAE}$, \ie, the zero-diagonal constraints.
\begin{equation}\label{eq:ease_diagonal2}
\hat{\mathbf{B}}_{EASE^{R}} - \hat{\mathbf{B}}_{LAE} = - \mathbf{V}\text{diag}\left( \frac{1}{\sigma_{1}^{2}+\lambda}, \ldots ,\frac{1}{\sigma_{n}^{2}+\lambda} \right)\mathbf{V}^{\top} \text{diagMat}(\boldsymbol{\mu})
\end{equation}

Note that the product of an arbitrary matrix $\mathbf{A}$ and a diagonal matrix $\mathbf{D}$ can be calculated by column-wise scaling effect on $\mathbf{A}$ for the corresponding diagonal entry of $\mathbf{D}$, \ie, ${\left( \mathbf{A} \cdot \mathbf{D} \right)}_{jj}=\mathbf{A}_{*j}\mathbf{D}_{jj}$. Thus, we focus on analyzing $-\mathbf{V} {\left( \left(\boldsymbol{\Sigma}^{\top} \boldsymbol{\Sigma}\right) + \lambda \mathbf{I} \right)}^{-1} \mathbf{V}^{\top}$ since the Lagrangian multipliers $\boldsymbol{\mu}$ only serve to adjust the coefficients.

\begin{figure}[t!]
\centering
\begin{tabular}{cc}
\includegraphics[width=0.223\textwidth]{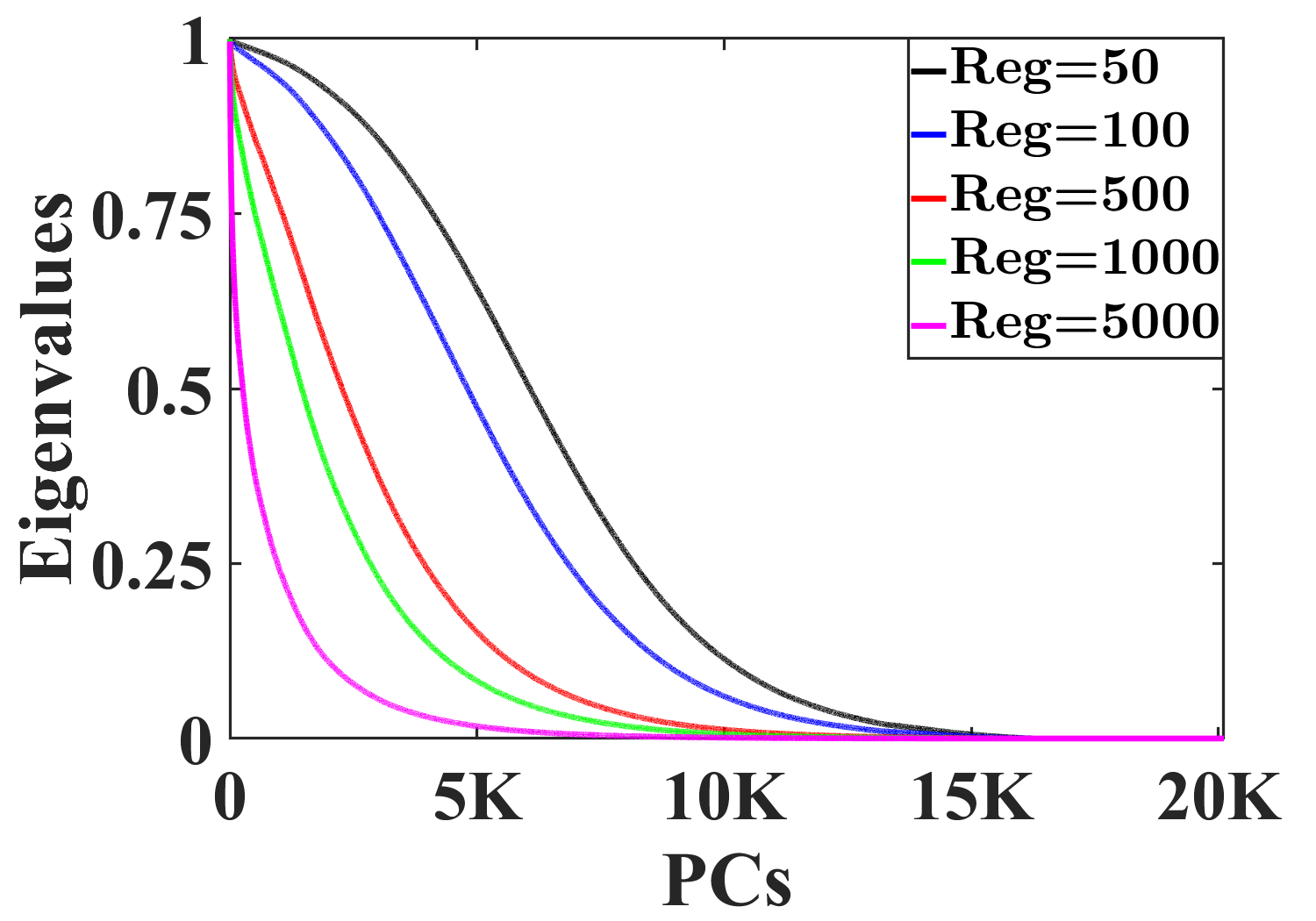} &
\includegraphics[width=0.223\textwidth]{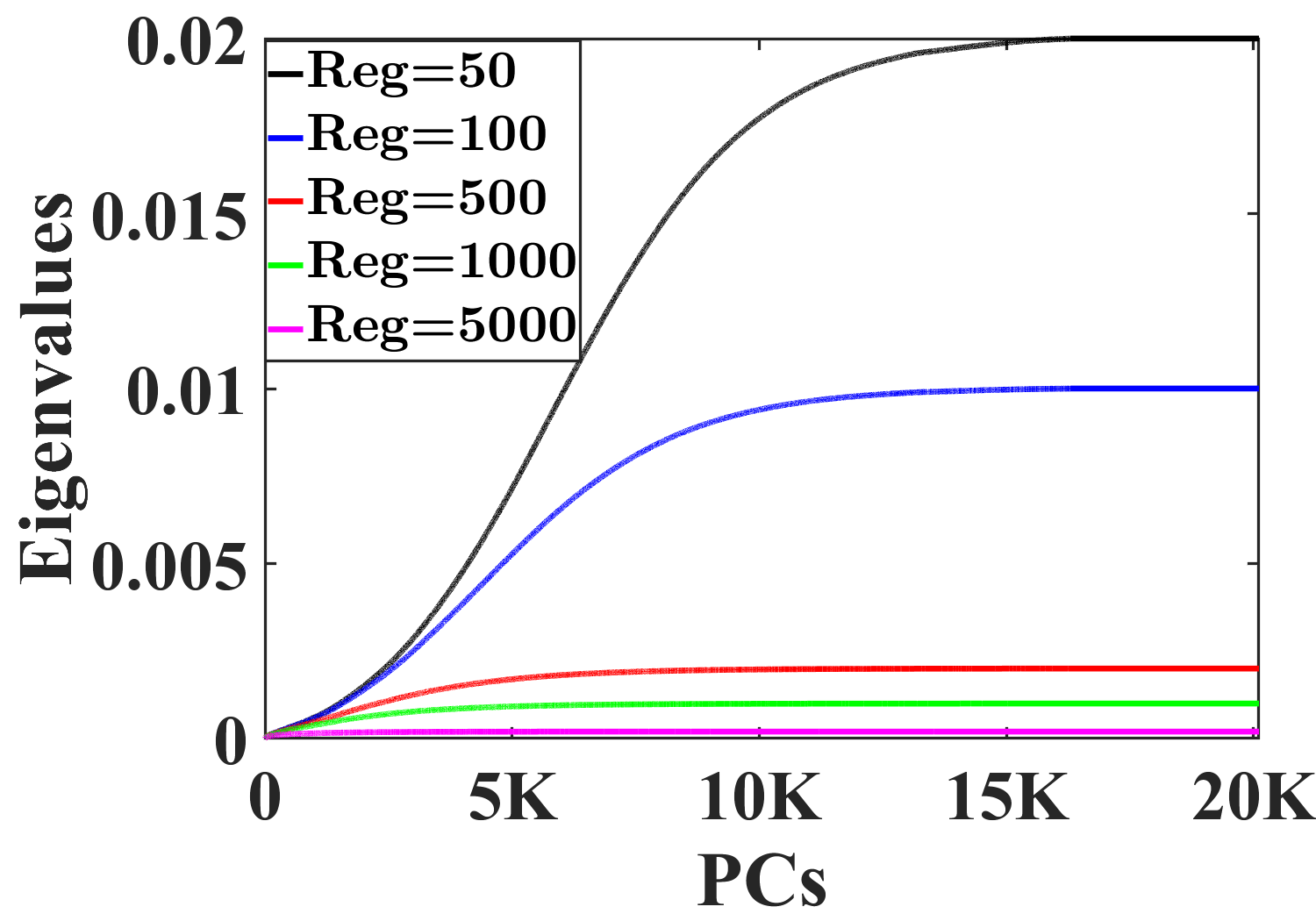} \\
(a) L2 regularization & (b) Diagonal constraints \\
\end{tabular}
\vspace{-2mm}
\caption{Distribution of eigenvalues scaled by (a) the L2 regularization $\left(\boldsymbol{\Sigma}^{\top}\boldsymbol{\Sigma} + \lambda \mathbf{I}\right)^{-1}\left(\boldsymbol{\Sigma}^{\top}\boldsymbol{\Sigma}\right)$ and (b) the zero-diagonal constraints $\left(\boldsymbol{\Sigma}^{\top}\boldsymbol{\Sigma} + \lambda \mathbf{I}\right)^{-1}$ on the ML-20M dataset with various $\lambda$.}
\vspace{-2mm}
\label{fig:eigenvalue_distribution_ml_20m}
\end{figure}

We then compare the eigenvalues of two terms: the L2 regularization and the zero-diagonal constraints. Figure~\ref{fig:eigenvalue_distribution_ml_20m} depicts the distribution of eigenvalues scaled by each term on the ML-20M dataset. The former is represented by the function of $\sigma_{i}^{2} / (\sigma_{i}^{2}+\lambda)$, and the latter is represented by $1 / (\sigma_{i}^{2}+\lambda)$ depending on $\sigma_{i}^{2}$ and $\lambda$. In Figure~\ref{fig:eigenvalue_distribution_ml_20m}(a), the eigenvalue of the L2 regularization ranges from $[0, 1]$. Assuming that $\sigma_{1}^{2}>\cdots>\sigma_{n}^{2}$, we observe that (i) the L2 regularization tends to be biased toward high-ranked principal components (PCs). (ii) As $\lambda$ increases, this tendency is strengthened, implying that high-ranked PCs highly influence $\hat{\mathbf{B}}_{LAE}$. We also find interesting observations in Figure~\ref{fig:eigenvalue_distribution_ml_20m}(b). In contrast to the L2 regularization, the eigenvalue of the diagonal constraints increases as $i$ goes high and ranges from $[0, 1 / \lambda]$. This observation shows (i) the zero-diagonal constraints tend to emphasize low-ranked PCs. Since it has a negative sign, it penalizes the low-ranked PCs in the solution. (ii) As $\lambda$ increases, this tendency weakens, implying that the impact of L2 regularization dominates the zero-diagonal constraint. In other words, the gap between $\hat{\mathbf{B}}_{LAE}$ and $\hat{\mathbf{B}}_{EASE^R}$ diminishes as $\lambda$ increases.

Our analyses can be extended to DLAE and EDLAE. The regularization term is replaced by $\text{diag}(\boldsymbol{\Lambda})=(\lambda_{1},\ldots,\lambda_{n})$, where $\lambda_{i} \ne \lambda_{j}$. By following the same procedure as Eqs.~\eqref{eq:ridge_svd} and~\eqref{eq:diagonal_svd}, $\hat{\mathbf{B}}_{DLAE}$ and the zero-diagonal constraints in $\hat{\mathbf{B}}_{EDLAE}$ can also be rewritten as follows, respectively.
\begin{equation}
\label{eq:dlae_svd}
\begin{split}
\hat{\mathbf{B}}_{DLAE} & = \left(\mathbf{X}^{\top} \mathbf{X}+\boldsymbol{\Lambda}\right)^{-1}\left(\mathbf{X}^{\top} \mathbf{X}\right) \\
%               & = \left(\mathbf{V} \left(\boldsymbol{\Sigma}^{\top}\boldsymbol{\Sigma}\right)\mathbf{V}^{\top} + \boldsymbol{\Lambda}\right)^{-1}\left(\mathbf{V} \left(\boldsymbol{\Sigma}^{\top}\boldsymbol{\Sigma}\right) \mathbf{V}^{\top}\right) \\
%               & = \left(\mathbf{V} \left(\boldsymbol{\Sigma}^{\top}\boldsymbol{\Sigma}\right)\mathbf{V}^{\top} + \mathbf{V}\mathbf{V}^{\top} \boldsymbol{\Lambda} \mathbf{V}\mathbf{V}^{\top}\right)^{-1}\left(\mathbf{V} \left(\boldsymbol{\Sigma}^{\top}\boldsymbol{\Sigma}\right) \mathbf{V}^{\top}\right) \\
%               & = \left(\mathbf{V}\left(\left(\boldsymbol{\Sigma}^{\top}\boldsymbol{\Sigma}\right) + \mathbf{V}^{\top}\boldsymbol{\Lambda}\mathbf{ V}\right)\mathbf{V}^{\top}\right)^{-1}\left(\mathbf{V}\left(\boldsymbol{\Sigma}^{\top}\boldsymbol{\Sigma}\right)\mathbf{V}^{\top}\right) \\
%               & = \mathbf{V}\left(\left(\boldsymbol{\Sigma}^{\top}\boldsymbol{\Sigma}\right) + \mathbf{V}^{\top}\boldsymbol{\Lambda}\mathbf{ V}\right)^{-1}\mathbf{V}^{\top}\left(\mathbf{V}\left(\boldsymbol{\Sigma}^{\top}\boldsymbol{\Sigma}\right)\mathbf{V}^{\top}\right) \\
               & = \mathbf{V} \left(\left(\boldsymbol{\Sigma}^{\top}\boldsymbol{\Sigma}\right) + \mathbf{V}^{\top}\boldsymbol{\Lambda}\mathbf{V}\right)^{-1} \left(\boldsymbol{\Sigma}^{\top}\boldsymbol{\Sigma}\right) \mathbf{V}^{\top} \\
\end{split}
\end{equation}
\begin{equation}
\label{eq:edlae_constraint_effect_svd}
\begin{split}
\hat{\mathbf{B}}_{EDLAE} - \hat{\mathbf{B}}_{DLAE} & = -\left(\mathbf{X}^{\top} \mathbf{X}+\boldsymbol{\Lambda}\right)^{-1} \text{diagMat}\left(\boldsymbol{\mu}\right) \\
%                                            & = \left(\mathbf{V} \left(\boldsymbol{\Sigma}^{\top}\boldsymbol{\Sigma}\right)\mathbf{V}^{\top} + \boldsymbol{\Lambda}\right)^{-1} \text{diagMat}\left(\boldsymbol{\mu}\right) \\
 %                                           & =  \left(\mathbf{V} \left(\boldsymbol{\Sigma}^{\top}\boldsymbol{\Sigma}\right)\mathbf{V}^{\top} + \mathbf{V}\mathbf{V}^{\top} \boldsymbol{\Lambda} \mathbf{V}\mathbf{V}^{\top}\right)^{-1} \text{diagMat}\left(\boldsymbol{\mu}\right) \\
  %                                          & = \left(\mathbf{V}\left(\left(\boldsymbol{\Sigma}^{\top}\boldsymbol{\Sigma}\right) + \mathbf{V}^{\top}\boldsymbol{\Lambda}\mathbf{V}\right)\mathbf{V}^{\top}\right)^{-1} \text{diagMat}\left(\boldsymbol{\mu}\right) \\
                                            & = -\mathbf{V}\left(\left(\boldsymbol{\Sigma}^{\top}\boldsymbol{\Sigma}\right) + \mathbf{V}^{\top}\boldsymbol{\Lambda}\mathbf{ V}\right)^{-1}\mathbf{V}^{\top} \text{diagMat}\left(\boldsymbol{\mu}\right) \\
\end{split}
\end{equation}

Although the forms are similar to LAE and EASE$^R$, we cannot further simplify $\mathbf{V}^{\top} \boldsymbol{\Lambda} \mathbf{V}$ in $\left(\left(\boldsymbol{\Sigma}^{\top} \boldsymbol{\Sigma}\right) + \mathbf{V}^{\top} \boldsymbol{\Lambda} \mathbf{V}\right)^{-1}$ because the commutative law between $\boldsymbol{\Lambda}$ and $\mathbf{V}$ does not always hold, $\mathbf{V}^{\top} \boldsymbol{\Lambda} \mathbf{V} \ne \boldsymbol{\Lambda}$. Thus, it is non-trivial to analyze DLAE and EDLAE with different regularization values in $\boldsymbol{\Lambda}$. Note that our analyses differ from those of the low-rank regression models in \cite{JinLGLCZ21KDD}. (We claim the formula expansions in~\cite{JinLGLCZ21KDD} are incorrect.)

\begin{figure}[t!]
\centering
\begin{tabular}{cc}
\includegraphics[width=0.228\textwidth]{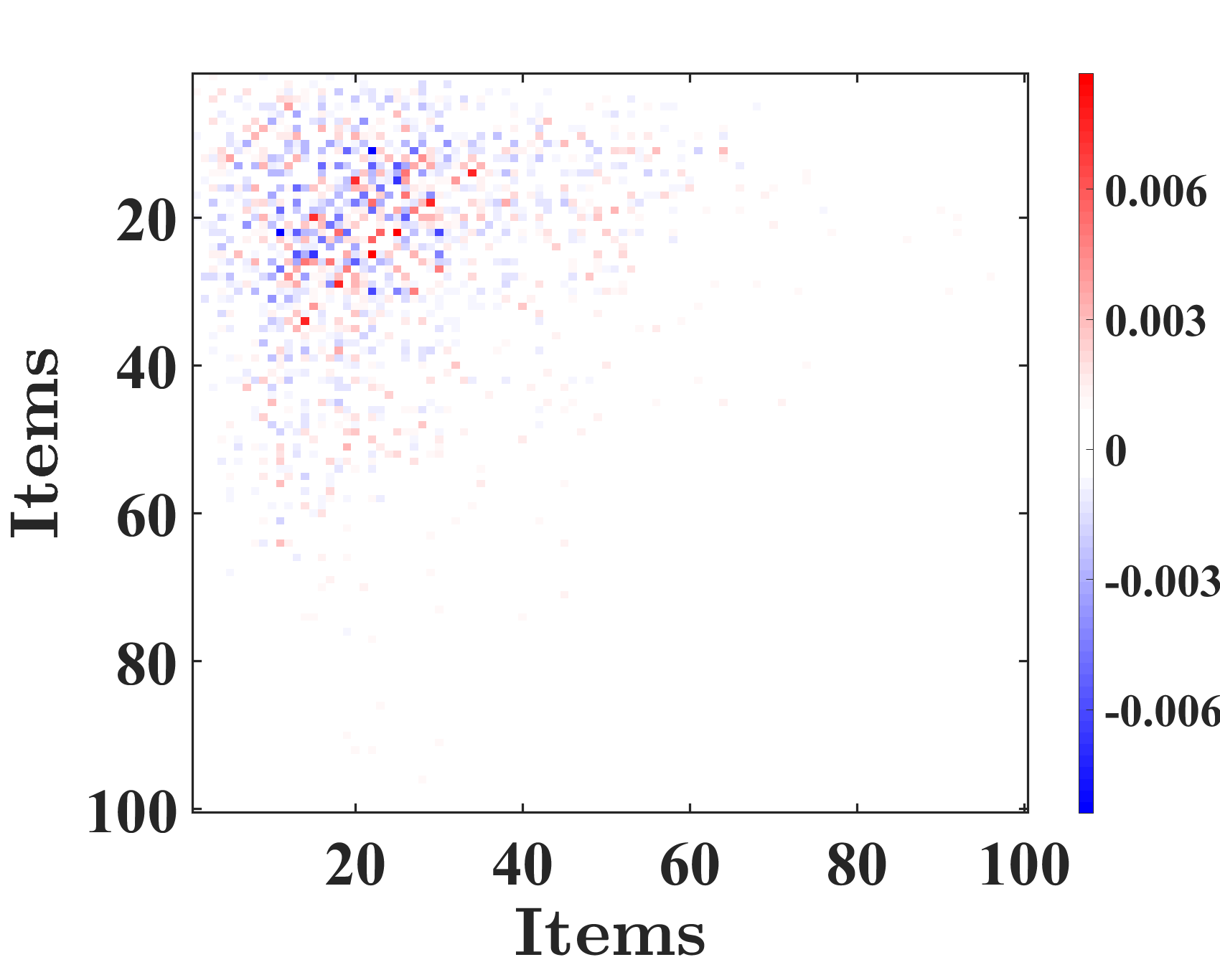} &
\includegraphics[width=0.228\textwidth]{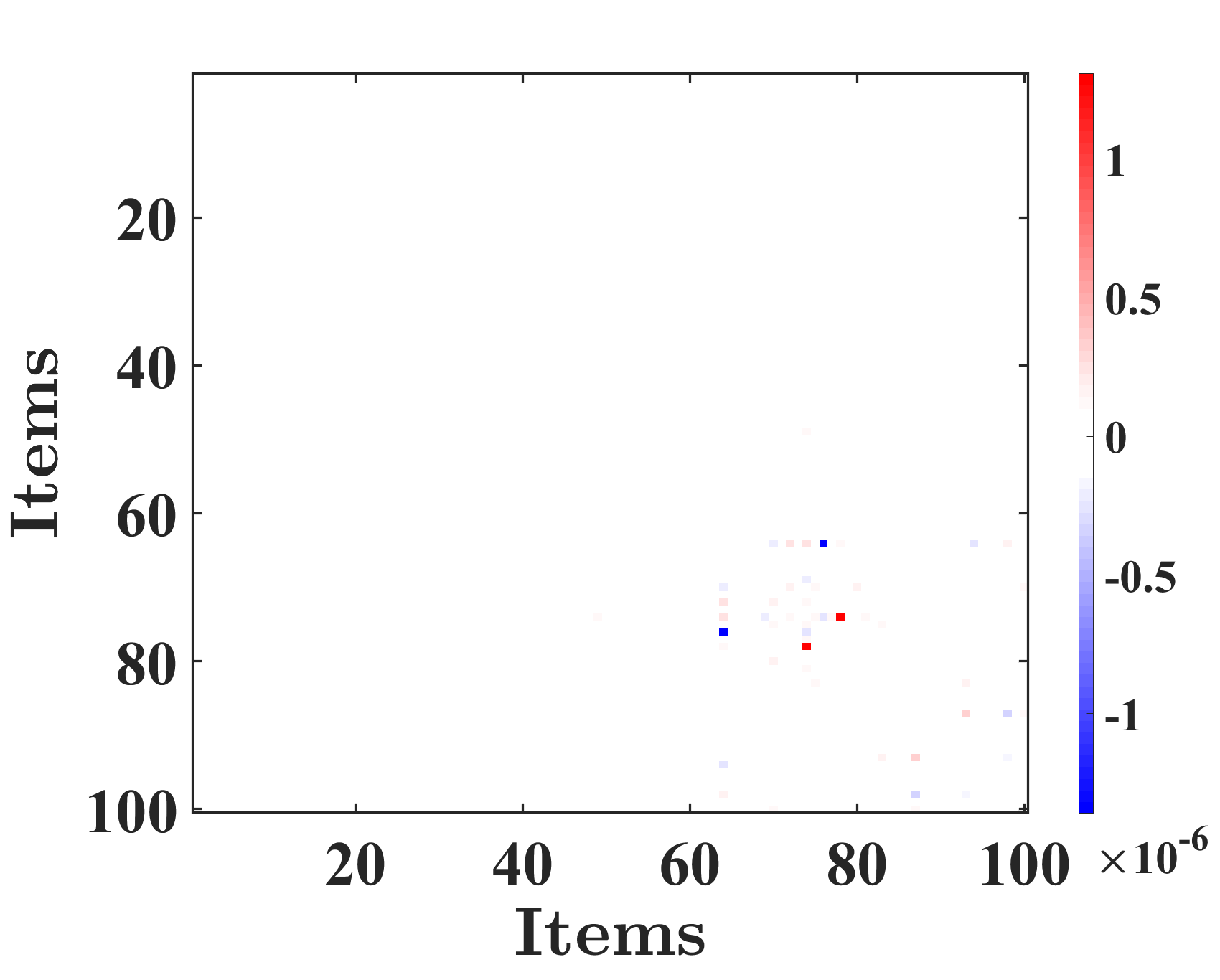} \\
\includegraphics[width=0.228\textwidth]{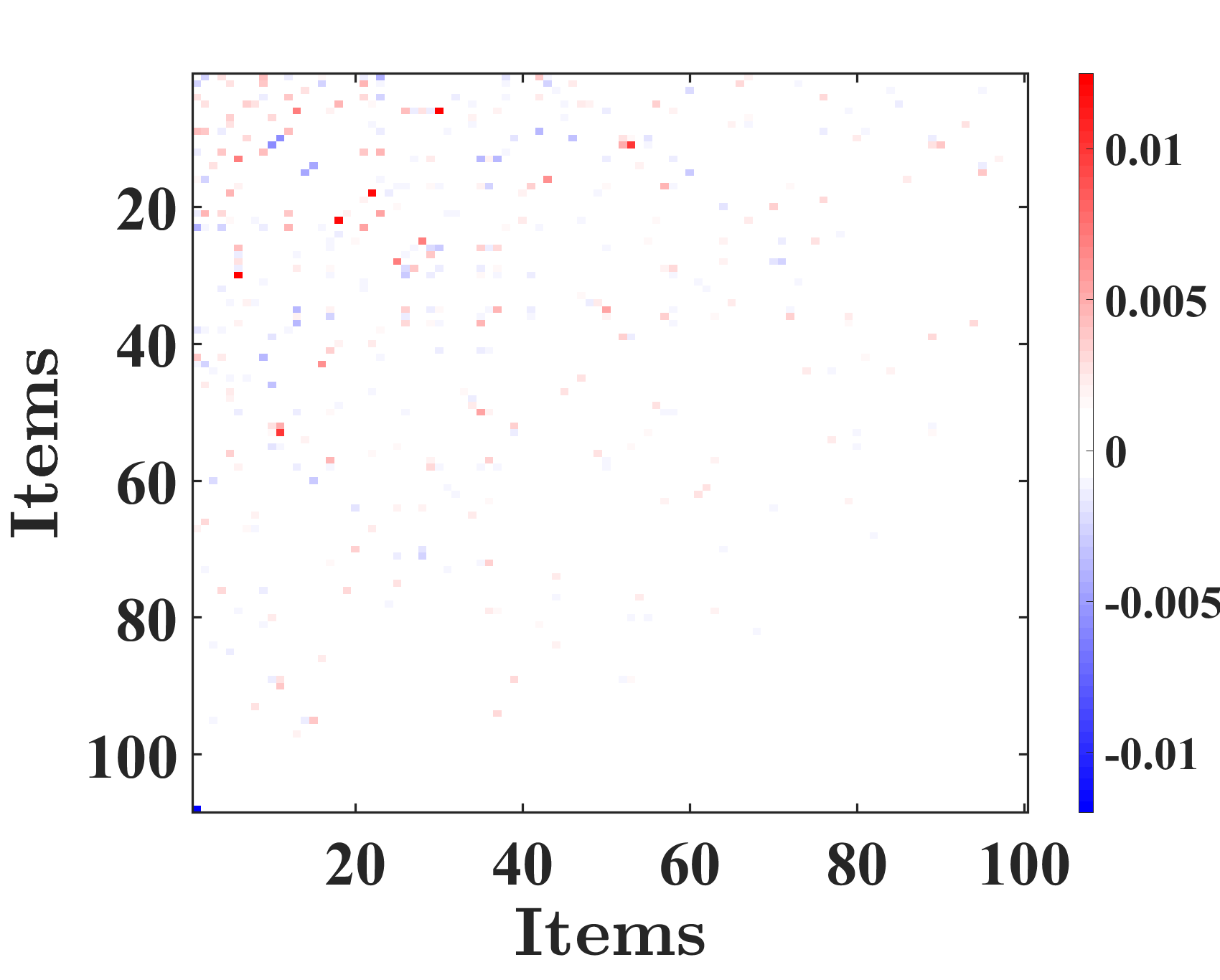} &
\includegraphics[width=0.228\textwidth]{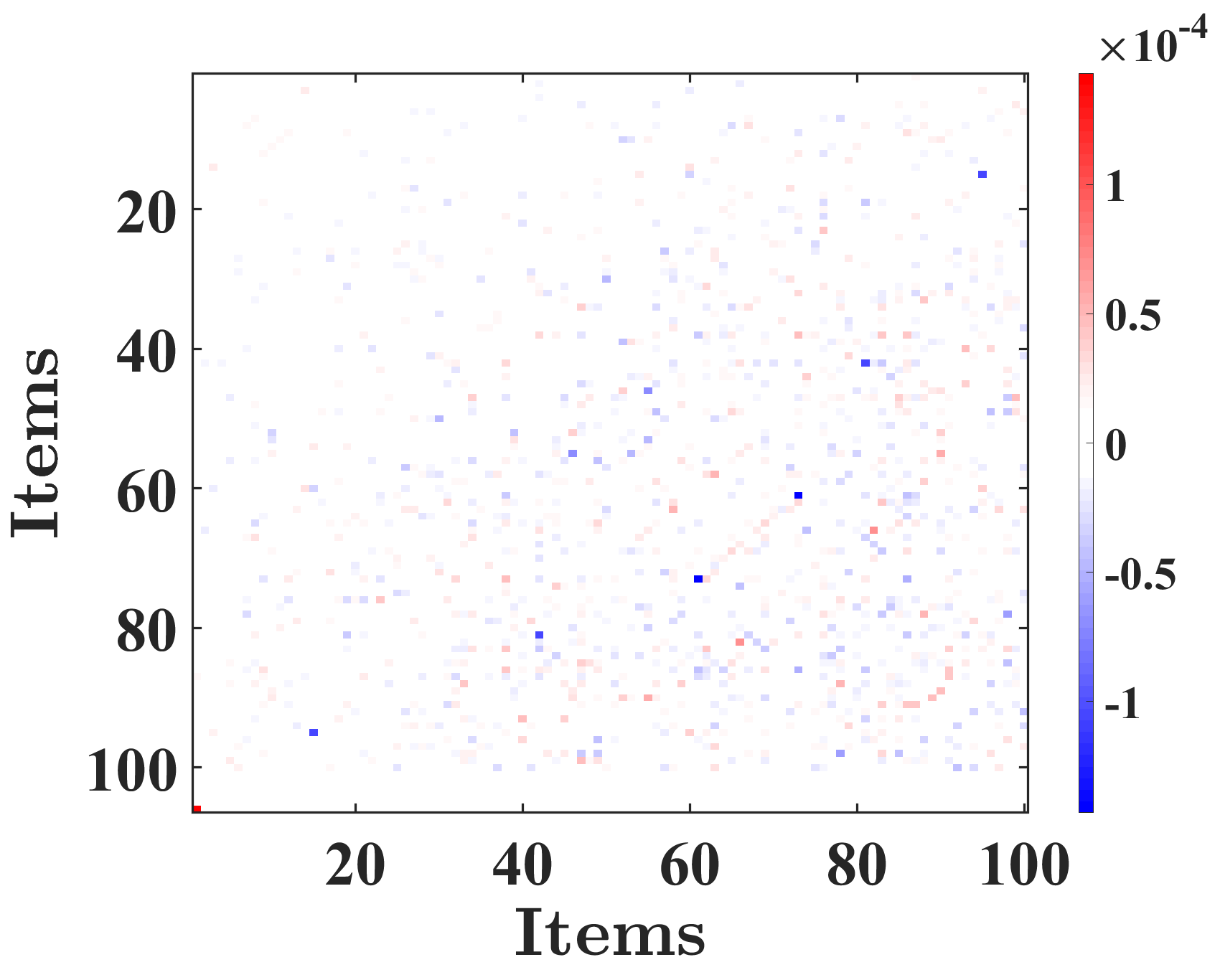} \\
(a) High-ranked PCs & (b) Low-ranked PCs \\
\end{tabular}
\vspace{-2mm}
\caption{Visualization of (a) high-ranked PCs and (b) low-ranked PCs of LAE solution, on two datasets (Top: ML-20M, Bottom: Yelp2018). Among 100 items, the first 20 items are popular, and the remaining 80 are unpopular. Items are sorted by popularity.}
\label{fig:PCs_visualization_ml-20m_yelp2018}
\end{figure}

We further analyze the effects of L2 regularization and diagonal constraints on recommendations in terms of item popularity. Since the gram matrix is represented by an eigenvalue decomposition form, we utilize principal component analysis (PCA). Assuming that $\sigma_{1}^{2}>\cdots>\sigma_{n}^{2}$, the high-ranked PCs contain strong collaborative signals. Since popular items have more chances to co-occur, they are closely related to high-ranked PCs. Meanwhile, low-ranked PCs capture weak collaborative signals.

We conduct a pilot study to better understand the difference between two PC groups (high vs. low) by analyzing Eq.~\eqref{eq:lae_svd}. Figure~\ref{fig:PCs_visualization_ml-20m_yelp2018}(a) and (b) visualize the high-ranked and low-ranked PCs, respectively, on the ML-20M and Yelp2018 datasets. We choose the top 20\% and bottom 20\% of PCs as two groups and then aggregate them with a weighted sum, using their corresponding eigenvalues as coefficients. For simplicity, we visualize only 100 items randomly drawn from each group, \ie, 20 for popular and 80 for unpopular groups.

On the ML-20M dataset, high-ranked PCs represent the collaborative signals between popular items, while low-ranked PCs represent the correlation between unpopular items (top of Figure~\ref{fig:PCs_visualization_ml-20m_yelp2018}). In other words, item popularity is highly related to the order of PCs regardless of the magnitude of eigenvalues. However, the trends on Yelp2018 are quite different from ML-20M (bottom of Figure~\ref{fig:PCs_visualization_ml-20m_yelp2018}). Since the item popularity bias of Yelp2018 is relatively lower than that of ML-20M, both high- and low-ranked PCs have a spread collaborative signal. Therefore, low-ranked PCs are likely to represent weak yet informative collaborative signals.

Based on this analysis, we discuss the effects of L2 regularization and zero-diagonal constraints. (i) L2 regularization mainly considers high-ranked PCs, implying that it removes weak collaborative signals. As $\lambda$ increases, it captures only strong collaborative signals and tends to recommend popular items. Notably, huge $\lambda$ provides only popularity-based recommendation results when the dataset is highly skewed to popular items. (ii) The zero-diagonal constraints mostly penalize low-ranked PCs, eliminating the weak collaborative signals (for the modest $\lambda$). When popular items dominate collaborative signals, using zero-diagonal constraints reduces weak collaborative signals and interrupts the recommendation of less popular items. As a result, it hinders the performance of long-tail recommendations.

\section{Relaxing Diagonal Constraints}\label{sec:model}

In this section, we propose novel linear autoencoder models via diagonal constraints relaxation, called \emph{Relaxed Linear AutoEncoder} (RLAE). First, we formulate the convex optimization problem of RLAE using diagonal inequality constraints and derive the solution of RLAE. We then analyze the relationship between RLAE and other linear autoencoder models, \ie, LAE and EASE$^{R}$. Our mathematical analysis shows that RLAE is a generalized version of linear autoencoder models with diagonal constraints.

\subsection{Convex Optimization Problem}
The convex optimization problem of RLAE is formulated by relaxing the diagonal constraints in Eq.~\eqref{eq:ease_objective}.
\begin{equation}\label{eq:model_objective}
    \min_{\mathbf{B}} \|\mathbf{X}-\mathbf{X}\mathbf{B}\|_F^2 + \lambda \|\mathbf{B}\|_F^2 \ \ s.t. \ \ \text{diag}(\mathbf{B}) \le \xi,
\end{equation}
where $\xi$ is the hyperparameter for relaxing diagonal constraints. When $\xi = 0$, it is equivalent to zero-equality constraints, implying EASE$^{R}$. If there are no diagonal constraints, \ie, $\xi \ge 1$, then RLAE is induced to LAE. (In Section~\ref{sec:proof}, we prove the relation of RLAE to other linear models by controlling the $\xi$.)

% The solution can be determined by minimizing Lagrangian formed by the convex optimization problem~\eqref{eq:model_objective} with respect to $\textbf{B}$.

% \begin{equation}\label{eq:model_Lagrangian}
%     L = \|\mathbf{X}-\mathbf{X}\mathbf{B}\|_F^2 + \lambda \|\mathbf{B}\|_F^2 + \boldsymbol{\mu}^{\top} \text{diag}(\mathbf{B} - \xi)
% \end{equation}

RLAE still achieves a closed-form solution via Karush–Kuhn–Tucker (KKT) conditions. Surprisingly, the solution form of RLAE is equivalent to that of EASE$^{R}$.
\begin{equation}\label{eq:model_solution}
\begin{split}
    \hat{\mathbf{B}}_{RLAE} & = \left(\mathbf{X}^{\top} \mathbf{X} + \lambda \mathbf{I}\right)^{-1} \left(\mathbf{X}^{\top} \mathbf{X} - \text{diagMat} (\boldsymbol{\mu})\right) \\
                   & = \mathbf{I} - \mathbf{P} \cdot \text{diagMat}(\lambda + \boldsymbol{\mu}),
    \end{split}
\end{equation}
\begin{equation}
  \label{eq:model_lagrangian}
  \begin{aligned}
    \text{where} \ {\mu}_{j} = &
    \begin{cases}
        \ \ \ 0 & \text{if} \ \ \ 1 - \mathbf{P}_{jj} \lambda \le \xi, \\
        \ \ \ \frac{1-\xi}{\mathbf{P}_{jj}} - \lambda & \text{otherwise}. 
    \end{cases}
\end{aligned}
\end{equation}

The key difference is that the diagonal vector $\boldsymbol{\mu}$ is determined by the inequality condition, \ie, $1 - \mathbf{P}_{jj} \lambda \le \xi$. If the inequality condition is satisfied, it becomes 0, \ie, $\mu_j = 0$. Otherwise, $\mu_{j}$ is equal to $\frac{1-\xi}{\mathbf{P}_{jj}} - \lambda$, relaxing the equality constraints by $\xi$. That is, the diagonal constraints are controlled by $\xi$.

We can also modify the convex optimization problem of DLAE by applying the diagonal inequality constraints. We call it \emph{Relaxed Denoising Linear AutoEncoder} (RDLAE).
\begin{equation}\label{eq:RDLAE_objective}
\min_{\mathbf{B}} \|\mathbf{X} - \mathbf{X} \mathbf{B}\|_F^2 + \|\boldsymbol{\Lambda}^{1 / 2} \cdot \mathbf{B}\|_F^2 \ \ s.t. \ \ \text{diag}(\mathbf{B}) \le \xi, \\
\end{equation}
where $\boldsymbol{\Lambda} = \frac{p}{1-p} \cdot \text{diagMat}\left(\text{diag}(\mathbf{X}^{\top}\mathbf{X})\right) + \lambda \mathbf{I}$.

Also, the solution of RDLAE can be formulated by the closed-form equation. Let $\mathbf{P}' = \left(\mathbf{X}^{\top} \mathbf{X} + \boldsymbol{\Lambda}\right)^{-1}$, then the optimization problem Eq.~\eqref{eq:RDLAE_objective} yields the following solution.
\begin{equation}\label{eq:RDLAE_solution}
\begin{split}
    \hat{\mathbf{B}}_{RDLAE} & = \left(\mathbf{X}^{\top} \mathbf{X} + \boldsymbol{\Lambda}\right)^{-1}\left(\mathbf{X}^{\top} \mathbf{X} -\text{diagMat}(\boldsymbol{\mu})\right) \\
                   & = \mathbf{I} - \mathbf{P}' \cdot \text{diagMat}\left(\text{diag}(\Lambda) + \boldsymbol{\mu}\right),
    \end{split}
\end{equation}
\begin{equation}
  \label{eq:RDLAE_lagrangian}
  \begin{aligned}
    \text{where} \ {\mu}_{j} = &
    \begin{cases}
        \ \ \ 0 & \text{if} \ \ \ 1 - \mathbf{P}'_{jj} \Lambda_{j} \le \xi, \\
        \ \ \ \frac{1-\xi}{\mathbf{P}'_{jj}} - \Lambda_{j} & \text{otherwise}. 
    \end{cases}
\end{aligned}
\end{equation}

\subsection{Upper and Lower Bounds Analysis of $\xi$}\label{sec:proof}

In this subsection, we show that RDLAE is a generalized version of two linear models, DLAE and EDLAE. We prove the upper and lower bounds of $\xi$. Note that our proof is also used for RLAE since it is a special case without using dropout, \ie, $p=0$.
\begin{theorem}\label{theorem:1}
If $\xi \ge 1$, the solution of RDLAE is equivalent to that of DLAE.
\begin{proof}
The condition for $\mu_j$ to be zero is as follows.
\begin{equation}
\label{eq:RDLAE_unconstrain_condition}
1 - \mathbf{P}'_{jj} \boldsymbol{\Lambda}_{jj} \le \xi \ \Leftrightarrow \	
\mathbf{P}'_{jj} \boldsymbol{\Lambda}_{jj} \ge 1 - \xi
\end{equation}

We will show that all diagonal entries of $\mathbf{P}'\boldsymbol{\Lambda}$ must be greater than or equal to zero. If a real-valued matrix is positive definite, then all diagonal entries are positive~\cite{H98matrix}. $(\mathbf{X}^{\top}\mathbf{X} + \boldsymbol{\Lambda})$ is positive definite because all diagonal entries of $\boldsymbol{\Lambda}$ are positive~\cite{H98matrix, HoerlK00Ridge, Steck20edlae}. Since the inverse of the positive definite matrix is a positive definite matrix, $\mathbf{P}'=(\mathbf{X}^{\top}\mathbf{X} + \boldsymbol{\Lambda})^{-1}$ is also positive definite. Therefore, the diagonal entries $\mathbf{P}'_{jj}$ are greater than zero for all $j$. Accordingly, if $\xi \ge 1$, the condition~\eqref{eq:RDLAE_unconstrain_condition} is satisfied for all $j$, and thus the Lagrangian multiplier vector $\boldsymbol{\mu}$ becomes a zero vector. As a result, the diagonal constraints of RDLAE are ignored, indicating the solution of DLAE.
\end{proof}
\end{theorem}

\begin{theorem}\label{theorem:2}
If $\xi = 0$, the solution of RDLAE is equivalent to that of EDLAE.
\begin{proof}
In the solution of RDLAE, the condition for the $j$th diagonal entry to be constrained is as follows.
\begin{equation}
\label{eq:RDLAE_constrain_condition}
1 - \mathbf{P}'_{jj} \boldsymbol{\Lambda}_{jj} \ge \xi \ \Leftrightarrow \ 
\mathbf{P}'_{jj} \boldsymbol{\Lambda}_{jj} \le 1 - \xi
\end{equation}

We will show that all diagonal entries of $\mathbf{P}'\boldsymbol{\Lambda}={(\mathbf{X}^{\top}\mathbf{X} + \boldsymbol{\Lambda})^{-1}}\boldsymbol{\Lambda}$ must be less than 1. $\mathbf{X}^{\top}\mathbf{X}$ is positive semi-definite, so all diagonal entries are semi-positive, \ie, greater or equal to zero~\cite{H98matrix}. From the proof of Theorem~\ref{theorem:1}, $\mathbf{P}'=(\mathbf{X}^{\top}\mathbf{X} + \boldsymbol{\Lambda})^{-1}$ is positive definite. Since the matrix multiplication between positive and positive semi-definite matrices yields a positive semi-definite matrix~\cite{H98matrix}, $(\mathbf{X}^{\top}\mathbf{X} + \boldsymbol{\Lambda})^{-1}(\mathbf{X}^{\top}\mathbf{X}) = \mathbf{I} - (\mathbf{X}^{\top}\mathbf{X} + \boldsymbol{\Lambda)}^{-1}\boldsymbol{\Lambda} = \mathbf{I} - \mathbf{P}'\boldsymbol{\Lambda}$ is positive semi-definite. Since the diagonal entries of $\mathbf{I} - \mathbf{P}'\boldsymbol{\Lambda}$ are all semi-positive, $\mathbf{P}'_{jj} \boldsymbol{\Lambda}_{jj}$ is less or equal to 1 for all $j$. In other words, the condition~\eqref{eq:RDLAE_constrain_condition} is satisfied for all $j$. As a result, all diagonal entries of RDLAE are constrained to be zero, indicating the solution of EDLAE.
\end{proof}
\end{theorem}

\section{Experimental Setup}\label{sec:setup}

\begin{table}[t] \small
\begin{center}
\caption{Statistics of six benchmark datasets: ML-20M, Netflix, MSD, Gowalla, Yelp2018, and Amazon-book.}
\vspace{-2mm}
\label{tab:statistics}
\begin{tabular}{l|rrrrr}
\toprule
Dataset     & \#Users   & \#Items   & \#Ratings     & Density   & $Gini_{item}$ \\
\hline
ML-20M      & 136,677   & 20,108    & 10.0M         & 0.36\%    & 0.90          \\
Netflix     & 463,435   & 17,769    & 56.9M         & 0.69\%    & 0.86          \\%& 40,000 \\
MSD         & 571,355   & 41,140    & 33.6M         & 0.36\%    & 0.56          \\% \\
\hline
Gowalla     & 29,858    & 40,981    & 1,027,370     & 0.01\%    & 0.44 \\
Yelp2018    & 31,668    & 38,048    & 1,561,406     & 0.13\%    & 0.51 \\
Amazon-book & 52,643    & 91,599    & 2,984,108     & 0.06\%    & 0.46 \\
\bottomrule
\end{tabular}
\end{center}
\vspace{-1mm}
\end{table}

% \vspace{1mm}
\noindent
\textbf{Datasets}. We extensively conduct experiments and analyses on six benchmark datasets, such as ML-20M, Netflix, MSD, Gowalla, Yelp2018, and Amazon-book, widely used in existing studies~\cite{Wang0WFC19NGCF, 0001DWLZ020LightGCN, 0002JP21LTOCF, Steck19EASE, Steck20edlae, ChinCC22}. In~\cite{ChinCC22}, they are categorized into different dataset groups. (i) Gowalla, Yelp2018, and Amazon-book are mainly used to evaluate matrix factorization models due to the characteristics of a relatively small number of users and high sparsity. (ii) Meanwhile, ML-20M, Netflix, and MSD are usually used to evaluate autoencoder models due to a large number of users. For reproducibility, we follow the preprocessing used in~\cite{Wang0WFC19NGCF} and~\cite{LiangKHJ18MultVAE}. Table~\ref{tab:statistics} summarizes the statistics of the datasets. % For ML-20M and Netflix, we binarize ratings above four as positive feedback and remove the users with fewer than five interactions. For MSD, we remove users and items with fewer than 20 and 200 interactions, respectively. For Gowalla, Yelp2018, and Amazon-book, we use the 10-core setting, \ie, remove the user and item with fewer than ten interactions.

\noindent
\textbf{Baseline models}. We compare our models with state-of-the-art linear autoencoders, LAE, EASE$^R$~\cite{Steck19EASE}, DLAE, and EDLAE~\cite{Steck20edlae}. We also evaluate the following state-of-the-art CF models.
\begin{itemize}[leftmargin=5mm]
    \item GRMF~\cite{RaoYRD15GRMF} is the MF model that enhances the smoothness of embeddings via the graph Laplacian regularizer. Following \cite{0001DWLZ020LightGCN, 0002JP21LTOCF}, we used BPR loss for model training.
    \item MultVAE~\cite{LiangKHJ18MultVAE} is the neural autoencoder model using variational inference.
    \item LightGCN~\cite{0001DWLZ020LightGCN} is the neural MF model using simplified graph convolutional networks (GCNs).
    \item LT-OCF~\cite{0002JP21LTOCF} is the neural MF model using learnable-time graph convolutional networks (GCNs) in which neural ODE (NODE) is used to find the optimal number of GCN layers.
    \item GF-CF~\cite{ShenWZSZLL21GFCF} is the linear autoencoder model that combines normalized singular vectors with linear and ideal low-pass filters.
    \item HMLET~\cite{KongKJ0LPK22HMLET} is the neural MF model using linear and non-linear hybrid graph convolutional networks (GCNs) with gating modules.
\end{itemize}

% \vspace{0.5mm}
\noindent
\textbf{Evaluation protocols and metrics}: For extensive evaluations, we adopt two evaluation protocols~\cite{LiangKHJ18MultVAE, Wang0WFC19NGCF}. While existing CF models merely employ one of the two protocols, we validate the generalized efficacy of our models on both of them.
\begin{itemize}[leftmargin=5mm]    
    \item \textbf{Strong generalization}: It randomly holds out a set of 80\% users for the training set. The remaining half and the others are used for the validation and test set, respectively. We use weak generalization for the validation and test sets; assuming the user has 80\% own ratings, CF models provide top-$N$ recommendation lists for 20\% unseen ratings. Because it evaluates unseen users as the test set, it is more applicable to real-world scenarios.
    
    \item \textbf{Weak generalization}. According to the conventional protocol~\cite{Wang0WFC19NGCF, 0001DWLZ020LightGCN, 0002JP21LTOCF}, we randomly split a user-item interaction matrix into 80\% training and 20\% test matrices.
\end{itemize}

We use two evaluation metrics, \emph{Recall} and \emph{Normalized Discounted Cumulative Gain (NDCG)}, widely used in the literature~\cite{LiangKHJ18MultVAE, Wang0WFC19NGCF, 0001DWLZ020LightGCN, 0002JP21LTOCF}. While recall quantifies how many preferred items exist, NDCG accounts for the position of preferred items in the top-$N$ recommendation list. To further analyze linear autoencoders, we adopt \emph{Average-Over-All (AOA)} and \emph{unbiased} evaluation~\cite{YangCXWBE18Unbias}. The unbiased evaluation measures true relevance under the \emph{missing-not-at-random (MNAR)} assumption, so it helps mitigate the impact of popularity bias. We use $\gamma=2$ as the normalization parameter for unbiased evaluation, which is commonly used in existing studies~\cite{YangCXWBE18Unbias, LeePLL22BISER}. We also report the results of two item groups, \ie, head and tail items, in terms of AOA evaluation. The head items are the top 20\% most popular, and the tail items are the rest.

\vspace{0.5mm}
\noindent
\textbf{Reproducibility}. We reproduced the experimental results of non-linear baselines using the hyperparameter settings provided in the original papers~\cite{0001DWLZ020LightGCN, 0002JP21LTOCF}. We conducted a grid search for linear autoencoder models to find optimal hyperparameters. The L2 regularization coefficient $\lambda$ was searched over $[1, 2, \ldots, 10, 20]$. We searched both the dropout probability $p$ and the inequality threshold $\xi$ in the range of $[0.1, 0.2, \ldots, 0.9]$. For GF-CF~\cite{ShenWZSZLL21GFCF}, $\alpha$ was searched in the range of $[0.0, 0.1, \ldots, 1.0]$. All the experiments were conducted on a desktop with 2 NVidia A6000, 512 GB memory, and 2 Intel Xeon Gold 6226R (2.90 GHz, 22.53M cache). Our implementations are available at \href{https://github.com/jaewan7599/RDLAE_SIGIR2023}{https://github.com/jaewan7599/RDLAE\_SIGIR2023}.

\section{Experimental Results}\label{sec:results}
In this section, we report the experimental results of two evaluation protocols, \ie, strong and weak generalization, on six benchmark datasets against theoretical discussions. Specifically, we address the following research questions:
\begin{itemize}[leftmargin=5mm]
    \item (RQ1) Do RLAE and RDLAE effectively provide recommendations that alleviate item popularity bias?
    \item (RQ2) How do regularization and diagonal constraints in linear autoencoders affect recommendations?
    \item (RQ3) How much do hyperparameters affect the performance of linear autoencoder models?
\end{itemize}

\begin{table*}[t]
\small
\caption{Performance comparison for the proposed methods and other linear autoencoder models on six datasets with the strong generalization protocol. The best results are marked in \red{bold}.}
\label{tab:strong_result}
\begin{center}
\begin{tabular}{c|c|cccc|cccc|cccc}
\toprule
\multirow{2}{*}{Dataset}    &   \multirow{2}{*}{Model}      & \multicolumn{4}{c|}{AOA}                                                      & \multicolumn{4}{c|}{Tail}                                                     & \multicolumn{4}{c}{Unbiased}                                                  \\
% ML-20M
\multirow{8}{*}{ML-20M}     
                            &                               & R@20              & N@20              & R@100             & N@100             & R@20              & N@20              & R@100             & N@100             & R@20              & N@20              & R@100             & N@100             \\  \hline
                            & GF-CF                         & 0.3250            & 0.2736            & 0.5765            & 0.3570            & 0.0029            & 0.0022            & 0.0111            & 0.0038            & 0.2188            & 0.0371            & 0.4334            & 0.0507            \\ 
                            & LAE                           & 0.3757            & 0.3228            & 0.6277            & 0.4070            & 0.0005            & 0.0001            & 0.0048            & 0.0011            & 0.2827            & 0.0473            & 0.5016            & 0.0606            \\
                            & EASE$^{R}$                    & 0.3905            & 0.3390            & 0.6363            & 0.4202            & 0.0052            & 0.0022            & 0.0215            & 0.0056            & 0.2857            & 0.0479            & 0.5113            & 0.0616            \\
                            & RLAE                          & \red{0.3913}      & \red{0.3402}      & \red{0.6394}      & \red{0.4224}      & \red{0.0137}      & \red{0.0069}      & \red{0.0579}      & \red{0.0167}      & \red{0.2951}      & \red{0.0487}      & \red{0.5283}      & \red{0.0626}      \\  \cline{2-14}
                            & DLAE                          & 0.3923            & 0.3408            & 0.6449            & 0.4241            & 0.0084            & 0.0047            & 0.0417            & 0.0120            & 0.2898            & 0.0477            & 0.5251            & 0.0620            \\
                            & EDLAE                         & 0.3925            & 0.3421            & 0.6410            & 0.4240            & 0.0066            & 0.0035            & 0.0269            & 0.0078            & 0.2859            & 0.0480            & 0.5128            & 0.0619            \\
                            & RDLAE                         & \red{0.3932}      & \red{0.3422}      & \red{0.6452}      & \red{0.4252}      & \red{0.0123}      & \red{0.0062}      & \red{0.0524}      & \red{0.0149}      & \red{0.2987}      & \red{0.0489}      & \red{0.5328}      & \red{0.0630}      \\  \hline
% Netflix
\multirow{7}{*}{Netflix}    
                            & GF-CF                         & 0.2972            & 0.2724            & 0.4973            & 0.3322            & 0.0185            & 0.0123            & 0.0468            & 0.0202            & 0.1868            & 0.0264            & 0.3563            & 0.0358            \\  
                            & LAE                           & 0.3465            & 0.3237            & 0.5410            & 0.3796            & 0.0066            & 0.0036            & 0.0258            & 0.0087            & 0.2357            & 0.0326            & 0.4068            & 0.0411            \\
                            & EASE$^{R}$                    & 0.3618            & 0.3388            & 0.5535            & 0.3938            & 0.0404            & 0.0222            & 0.1093            & 0.0408            & 0.2554            & 0.0351            & 0.4321            & 0.0435            \\
                            & RLAE                          & \red{0.3623}      & \red{0.3392}      & \red{0.5551}      & \red{0.3945}      & \red{0.0585}      & \red{0.0377}      & \red{0.1342}      & \red{0.0574}      & \red{0.2606}      & \red{0.0355}      & \red{0.4362}      & \red{0.0437}      \\  \cline{2-14}
                            & DLAE                          & 0.3621            & 0.3400            & 0.5557            & 0.3950            & \red{0.0597}      & \red{0.0381}      & \red{0.1320}      & \red{0.0575}      & 0.2549            & 0.0355            & 0.4302            & 0.0438            \\
                            & EDLAE                         & 0.3659            & 0.3428            & 0.5583            & 0.3978            & 0.0470            & 0.0279            & 0.1141            & 0.0461            & 0.2569            & 0.0358            & 0.4328            & 0.0441            \\
                            & RDLAE                         & \red{0.3661}      & \red{0.3431}      & \red{0.5588}      & \red{0.3982}      & 0.0545            & 0.0344            & 0.1228            & 0.0527            & \red{0.2598}      & \red{0.0360}      & \red{0.4350}      & \red{0.0443}      \\  \hline
% MSD
\multirow{7}{*}{MSD}        
                            & GF-CF                         & 0.2513            & 0.2457            & 0.4310            & 0.3111            & 0.1727            & 0.1331            & 0.2978            & 0.1695            & 0.2137            & 0.0282            & 0.3658            & 0.0351            \\  
                            & LAE                           & 0.2848            & 0.2740            & 0.4778            & 0.3448            & 0.1862            & 0.1234            & 0.3715            & 0.1765            & 0.2568            & 0.0320            & 0.4411            & 0.0401            \\
                            & EASE$^{R}$                    & \textbf{0.3338}   & \textbf{0.3261}   & \textbf{0.5074}   & \textbf{0.3899}   & 0.2504            & 0.1758            & 0.4109            & 0.2232            & 0.3019            & 0.0377            & 0.4663            & 0.0448            \\
                            & RLAE                          & \red{0.3338}      & \red{0.3261}      & \red{0.5074}      & \red{0.3899}      & \red{0.2507}      & \red{0.1767}      & \red{0.4110}      & \red{0.2240}      & \red{0.3021}      & \red{0.0378}      & \red{0.4664}      & \red{0.0449}      \\  \cline{2-14}
                            & DLAE                          & 0.3288            & 0.3208            & 0.5103            & 0.3873            & \red{0.2526}      & \red{0.1863}      & 0.4107            & \red{0.2325}      & 0.2993            & 0.0378            & 0.4666            & \textbf{0.0450}   \\
                            & EDLAE                         & 0.3336            & 0.3258            & \red{0.5124}      & 0.3913            & 0.2503            & 0.1782            & 0.4105            & 0.2253            & 0.3014            & 0.0378            & \red{0.4684}      & \textbf{0.0450}   \\
                            & RDLAE                         & \red{0.3341}      & \red{0.3265}      & 0.5110            & \red{0.3914}      & 0.2511            & 0.1784            & \red{0.4109}      & 0.2255            & \red{0.3022}      & \red{0.0379}      & 0.4680            & \red{0.0450}      \\   \hline
% Gowalla
\multirow{7}{*}{Gowalla}
                            & GF-CF                         & 0.2252            & 0.1660            & \red{0.4529}      & 0.2318            & 0.1151            & 0.0591            & 0.2962            & 0.1049            & 0.1734            & 0.0343            & 0.3864            & 0.0508            \\  
                            & LAE                           & 0.2271            & 0.1706            & 0.4491            & 0.2346            & 0.0799            & 0.0371            & 0.2680            & 0.0836            & 0.1672            & 0.0326            & 0.3763            & 0.0487            \\
                            & EASE$^{R}$                    & 0.2414            & 0.1831            & 0.4493            & 0.2437            & 0.0941            & 0.0428            & 0.2862            & 0.0909            & 0.1753            & 0.0335            & 0.3802            & 0.0491            \\
                            & RLAE                          & \red{0.2448}      & \red{0.1873}      & 0.4499            & \red{0.2468}      & \red{0.1243}      & \red{0.0625}      & \red{0.3177}      & \red{0.1113}      & \red{0.1912}      & \red{0.0370}      & \red{0.3922}      & \red{0.0522}      \\  \cline{2-14}
                            & DLAE                          & 0.2495            & 0.1891            & 0.4615            & 0.2507            & 0.1109            & 0.0532            & 0.3087            & 0.1026            & 0.1881            & 0.0366            & 0.3965            & 0.0524            \\
                            & EDLAE                         & 0.2469            & 0.1859            & \red{0.4619}      & 0.2484            & 0.0951            & 0.0432            & 0.2904            & 0.0918            & 0.1790            & 0.0344            & 0.3896            & 0.0504            \\
                            & RDLAE                         & \red{0.2499}      & \red{0.1900}      & 0.4594            & \red{0.2510}      & \red{0.1210}      & \red{0.0587}      & \red{0.3173}      & \red{0.1079}      & \red{0.1923}      & \red{0.0373}      & \red{0.3976}      & \red{0.0528}      \\  \hline
% Yelp2018
\multirow{7}{*}{Yelp2018}   
                            & GF-CF                         & 0.1134            & 0.0900            & \red{0.2858}      & 0.1487            & \red{0.0155}      & \red{0.0078}      & \red{0.0793}      & \red{0.0251}      & 0.0685            & 0.0081            & \red{0.2007}      & 0.0150            \\  
                            & LAE                           & 0.1160            & 0.0954            & 0.2720            & 0.1487            & 0.0086            & 0.0039            & 0.0635            & 0.0187            & 0.0705            & 0.0086            & 0.1914            & 0.0148            \\
                            & EASE$^{R}$                    & 0.1144            & 0.0933            & 0.2681            & 0.1458            & 0.0091            & 0.0042            & 0.0684            & 0.0201            & 0.0679            & 0.0081            & 0.1883            & 0.0143            \\
                            & RLAE                          & \red{0.1173}      & \red{0.0968}      & 0.2726            & \red{0.1499}      & 0.0127            & 0.0060            & 0.0784            & 0.0237            & \red{0.0735}      & \red{0.0089}      & 0.1972            & \red{0.0152}      \\  \cline{2-14}
                            & DLAE                          & \textbf{0.1190}   & 0.0971            & \red{0.2784}      & 0.1516            & 0.0121            & 0.0057            & 0.0773            & 0.0236            & 0.0724            & 0.0087            & 0.1986            & 0.0152            \\
                            & EDLAE                         & 0.1171            & 0.0957            & 0.2757            & 0.1499            & 0.0103            & 0.0049            & 0.0727            & 0.0219            & 0.0698            & 0.0084            & 0.1939            & 0.0147            \\
                            & RDLAE                         & \red{0.1190}      & \red{0.0976}      & 0.2773            & \red{0.1519}      & \red{0.0161}      & \red{0.0077}      & \red{0.0882}      & \red{0.0274}      & \red{0.0741}      & \red{0.0089}      & \red{0.2018}      & \red{0.0154}      \\  \hline
% Amazon-book
\multirow{7}{*}{Amazon-book}
                            & GF-CF                         & 0.1668            & 0.1492            & 0.3182            & 0.2009            & 0.0988            & \red{0.0702}      & 0.2051            & 0.1000            & 0.1401            & 0.0195            & 0.2740            & 0.0263            \\  
                            & LAE                           & 0.1920            & 0.1749            & 0.3239            & 0.2198            & 0.1012            & 0.0635            & 0.2218            & 0.0980            & 0.1644            & 0.0220            & 0.2896            & 0.0281            \\
                            & EASE$^{R}$                    & 0.1912            & 0.1734            & 0.3258            & 0.2193            & 0.0761            & 0.0444            & 0.1815            & 0.0746            & 0.1481            & 0.0195            & 0.2725            & 0.0256            \\
                            & RLAE                          & \red{0.1968}      & \red{0.1804}      & \red{0.3288}      & \red{0.2255}      & \red{0.1057}      & 0.0672            & \red{0.2260}      & \red{0.1018}      & \red{0.1649}      & \red{0.0221}      & \red{0.2909}      & \red{0.0282}      \\  \cline{2-14}
                            & DLAE                          & 0.1994            & 0.1820            & \red{0.3374}      & 0.2291            & 0.0993            & 0.0631            & 0.2187            & 0.0972            & 0.1637            & 0.0220            & \red{0.2938}      & 0.0283            \\
                            & EDLAE                         & 0.1940            & 0.1756            & 0.3349            & 0.2240            & 0.0829            & 0.0512            & 0.1938            & 0.0827            & 0.1523            & 0.0205            & 0.2820            & 0.0268            \\
                            & RDLAE                         & \red{0.2011}      & \red{0.1834}      & 0.3354            & \red{0.2293}      & \red{0.1043}      & \red{0.0670}      & \red{0.2214}      & \red{0.1007}      & \red{0.1663}      & \red{0.0225}      & 0.2934            & \red{0.0286}      \\
\bottomrule
\end{tabular}
\end{center}
\end{table*}

\subsection{Performance Comparison (RQ1, RQ2)}\label{sec:overall performance}
\textbf{Strong generalization}. We compare seven linear models, including GF-CF~\cite{ShenWZSZLL21GFCF}, to evaluate the effectiveness of our models. We also perform quantitative analyses of how the regularization and the diagonal constraints affect the performance of linear autoencoder models. Table~\ref{tab:strong_result} shows the performance of top-$K$ recommendation on six datasets. Note that we observe similar trends between head items and AOA metrics in the ML-20M, Netflix, and MSD datasets, so we do not report the performance for head items in Table~\ref{tab:strong_result}.

As shown in Table~\ref{tab:strong_result}, our models, \ie, RLAE and RDLAE, consistently outperform other linear autoencoder models in both AOA and unbiased evaluation on all six datasets, more significantly in the latter. RLAE and RDLAE show performance gains in the unbiased evaluation of 1.62\% and 1.61\% on NDCG@100 for the ML-20M dataset, respectively, and 2.70\% and 1.32\% for the Yelp2018 dataset, respectively. The significant enhancement in tail items drives the improvement in unbiased evaluation, indicating that they can adequately mitigate the popularity bias. Specifically, RLAE and RDLAE show performance gains of 198.21\% and 24.17\% in tail items on NDCG@100 for the ML-20M dataset, and 17.91\% and 16.10\% for the Yelp2018 dataset, respectively.

Our models outperform GF-CF, a state-of-the-art model, in almost all metrics. On the three datasets, ML-20M, Netflix, and MSD, RLAE outperforms GF-CF by an average of 20.80\% in AOA evaluation, 185.26\% in tail items, and 24.49\% in unbiased evaluation on NDCG@100. For the strong generalization datasets, using the precision matrix rather than the covariance matrix is better~\cite{Steck19EASE}. However, the linear filter, one of the main components of GF-CF, is similar to the concept of a covariance matrix, leading to lower performance of GF-CF. On the rest of the datasets, Gowalla, Yelp2018, and Amazon-book, RLAE outperforms GF-CF by an average of 6.47\% in AOA evaluation, 0.66\% in tail items, and 3.77\% in unbiased evaluation on NDCG@100. The ideal low-pass filter, another main component of GF-CF, emphasizes high-ranked PCs but needs to fully account for informative CF signals from low-ranked PCs in these datasets, causing RLAE to outperform GF-CF.

We observe that the popularity bias of the dataset has a significant impact on determining the utility of the diagonal constraint. On the datasets with large popularity bias, \ie, ML-20M, Netflix, and MSD datasets, EASE$^{R}$ outperforms LAE on all metrics, with an average performance gain of 268.17\% in tail items on NDCG@100. Here, the diagonal constraints help models focus on high-ranked PCs, preventing a large increase in L2 regularization. We find that the optimal $\lambda$ for LAE relative to EASE$^{R}$ is 12.5, 40, and 40 times larger for the three datasets, respectively. This makes unpopular items almost uninformative, resulting in the dramatic performance drop in tail items of LAE.

On the other hand, on the datasets with relatively low popularity bias, such as Gowalla, Yelp2018, and Amazon-book datasets, LAE performs slightly worse or even better than EASE$^{R}$. For these datasets, we observe that the optimal $\lambda$ tends to be lower, and the differences between the optimal $\lambda$s of the two models are reduced to a range of less than a factor of two or less. This low $\lambda$ takes into account the relatively uniform importance of PCs' rank and allows the model to make recommendations based on CF signals from low-ranked PCs as well as high-ranked PCs. In other words, low-ranked PCs are informative in these datasets, and the diagonal constraints that penalize them cause performance degradation. Specifically, LAE outperforms EASE$^{R}$ by 31.37\% in tail items on NDCG@100 for the Amazon-book dataset, even though LAE has a 25\% higher $\lambda$ than EASE$^{R}$.

\begin{table}
\small % \fontsize{9}{10} \selectfont
\caption{Performance comparison for linear autoencoder models and deep learning baseline models on Gowalla, Yelp2018, and Amazon-book datasets with the weak generalization protocol. The best results are marked in \red{bold}, and the second best models are \blue{underlined}.}
\label{tab:weak_result}
\begin{center}
\begin{tabular}{c|cc|cc|cc}
\toprule
Dataset         & \multicolumn{2}{c|}{Gowalla}          & \multicolumn{2}{c|}{Yelp2018}         & \multicolumn{2}{c}{Amazon-Book}       \\
Method          & R@20              & N@20              & R@20              & N@20              & R@20              & N@20              \\
\hline
GRMF            & 0.1477            & 0.1205            & 0.0571            & 0.0462            & 0.0354            & 0.0270            \\
GRMF-Norm       & 0.1557            & 0.1261            & 0.0561            & 0.0454            & 0.0352            & 0.0269            \\
Mult-VAE        & 0.1641            & 0.1335            & 0.0584            & 0.0450            & 0.0407            & 0.0315            \\
LightGCN        & 0.1830            & 0.1554            & 0.0649            & 0.0530            & 0.0411            & 0.0315            \\
LT-OCF          & \red{0.1875}      & \blue{0.1574}     & 0.0671            & 0.0549            & 0.0442            & 0.0341            \\
HMLET           & \blue{0.1874}     & \red{0.1589}      & 0.0675            & 0.0557            & 0.0482            & 0.0371            \\
\hline
GF-CF           & 0.1849            & 0.1536            & \red{0.0697}      & \red{0.0571}      & 0.0710            & 0.0584            \\
LAE             & 0.1630            & 0.1295            & 0.0658            & 0.0555            & 0.0746            & 0.0611            \\
EASE$^{R}$      & 0.1765            & 0.1467            & 0.0657            & 0.0552            & 0.0710            & 0.0566            \\
DLAE            & 0.1839            & 0.1533            & 0.0678            & \blue{0.0570}     & 0.0751            & 0.0610            \\
EDLAE           & 0.1844            & 0.1539            & 0.0673            & 0.0565            & 0.0711            & 0.0566            \\ \hline
RLAE            & 0.1772            & 0.1467            & 0.0667            & 0.0562            & \red{0.0754}      & \red{0.0615}      \\
RDLAE           & 0.1845            & 0.1539            & \blue{0.0679}     & 0.0569            & \red{0.0754}      & \blue{0.0613}     \\
\bottomrule
\end{tabular}
\end{center}
\end{table}

We find that dropout leads to an overall performance improvement, with the optimal $\lambda$ tending to decrease because dropout provides additional regularization. The lower $\lambda$ strengthens the model's concentration on low-ranked PCs, improving performance on unbiased evaluation and tail items. Specifically, on the ML-20M dataset, where the $\lambda$ of DLAE is reduced by 10 times compared to LAE's, DLAE significantly outperforms LAE by 900.91\% in tail items on NDCG@100, while on the Amazon-book dataset, where the reduction in $\lambda$ is relatively small, DLAE performs slightly worse than LAE with the performance degradation of 0.82\%.

Moreover, we observe that the tendency of the diagonal constraints is maintained regardless of whether dropout is applied, but its effectiveness is reduced. For example, on the Netflix and MSD datasets, EASE$^{R}$ outperforms LAE with a dramatic performance gain, while EDLAE is slightly better than DLAE. Furthermore, EDLAE performs slightly worse than DLAE on the ML-20M dataset. This is because dropout emphasizes high-ranked PCs similarly to diagonal constraints, and we discuss this in Section~\ref{sec:hyperparameter}.

\vspace{1mm}
\noindent
\textbf{Weak generalization}. Table~\ref{tab:weak_result} reports the results of weak generalization by comparing linear autoencoder and neural models. Note that existing studies rarely evaluate linear autoencoder models in this setting.

Our models, \ie, RLAE and RDLAE, still show higher or comparable overall performance compared to other linear autoencoder models in weak generalization. RDLAE is comparable to GF-CF on Gowalla, worse on Yelp2018, and considerably better on Amazon-book on Recall@20. Specifically, for the Amazon-book dataset, RDLAE outperforms GF-CF by 6.20\% and 5.31\% on Recall@20 and NDCG@20, respectively. In addition, reducing the diagonal constraints helps improve performance on datasets with a low popularity bias. For the Yelp2018 and the Amazon-book datasets, the models with zero-diagonal constraints, \ie, EASE$^{R}$ and EDLAE, show the worst performance. However, on Gowalla, introducing diagonal constraints offers better performance. Due to the high sparsity, low-ranked PCs in Gowalla may contain noisy information.

We highlight the comparison between RDLAE and HMLET~\cite{KongKJ0LPK22HMLET}, a state-of-the-art GCN model. HMLET performs better than RDLAE on the Gowalla dataset. However, as the dataset grows, the performance gain of RDLAE increases. Especially for the Amazon-Book dataset, the largest of the three datasets, RDLAE significantly outperforms HMLET with a 56.43\% gain on Recall@20. This is because most neural models are optimized for small datasets rather than large ones. The high sparsity of the dataset also hinders their performance. Meanwhile, linear autoencoders easily capture significant collaborative signals using the closed-form solution regardless of data sparsity.

\subsection{Hyperparameter Sensitivity Analysis (RQ3)}\label{sec:hyperparameter}
\begin{figure}[t!]
\centering
\begin{tabular}{cc}
\includegraphics[width=0.22\textwidth]{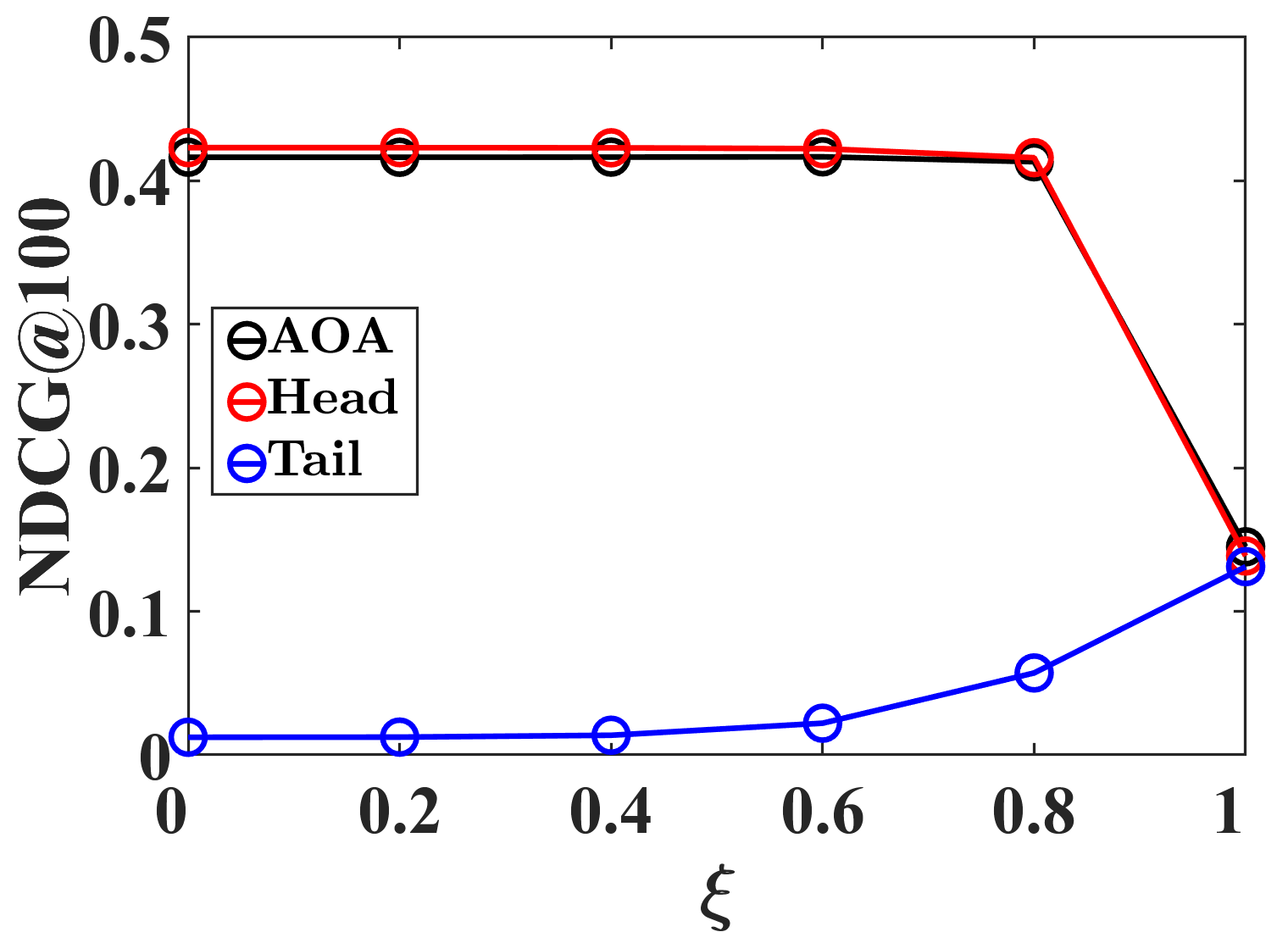} &
\includegraphics[width=0.22\textwidth]{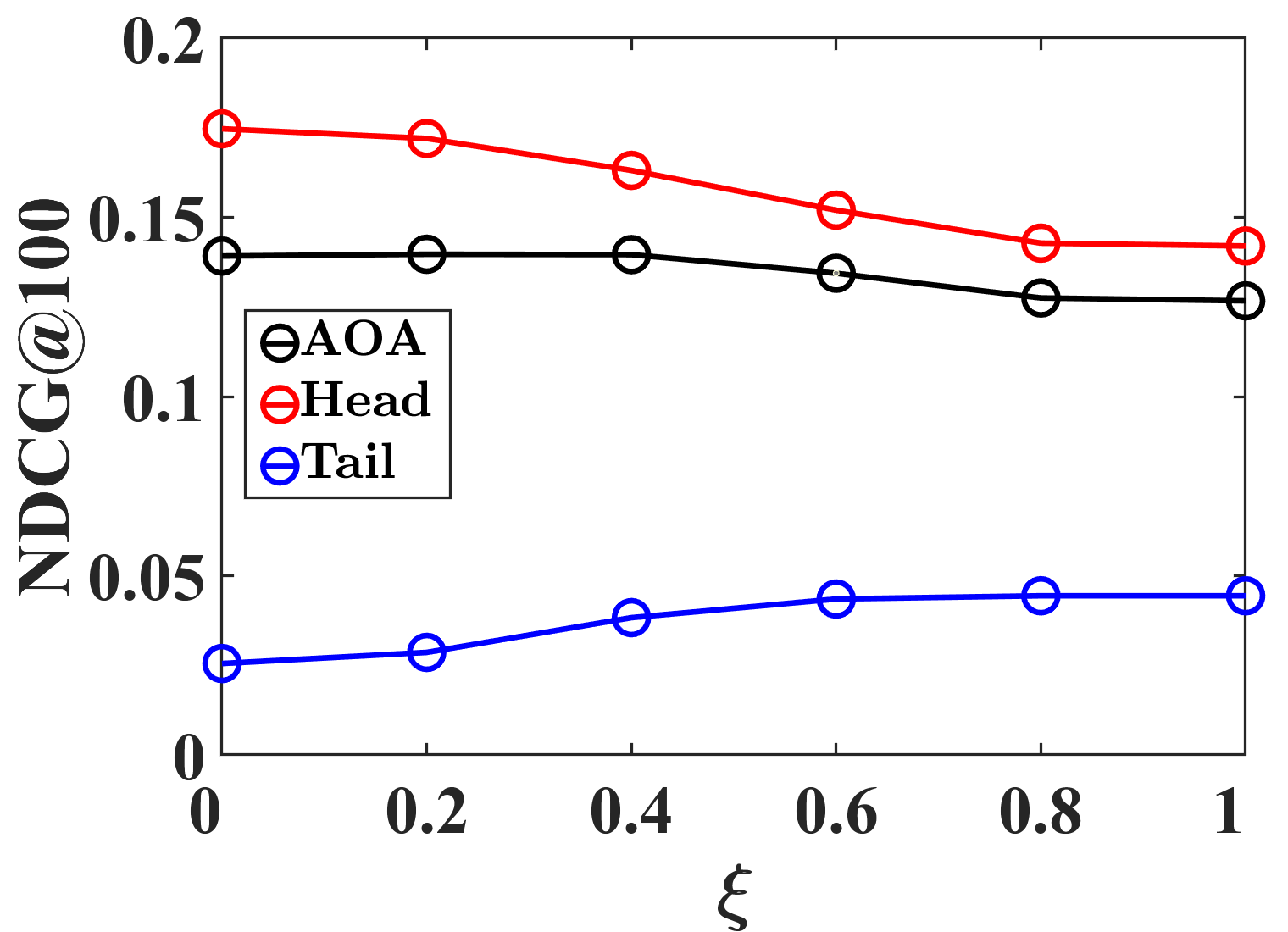} \\
(a) ML-20M & (b) Yelp2018 \\
\end{tabular}
\vspace{-2mm}
\caption{NDCG@100 of RLAE over various $\boldsymbol{\xi}$ for diagonal constraints on two datasets, ML-20M and Yelp2018.}
\vspace{-2mm}
\label{fig:various_xi}
\end{figure}

\begin{figure}[t!]
\centering
\begin{tabular}{cc}
\includegraphics[width=0.22\textwidth]{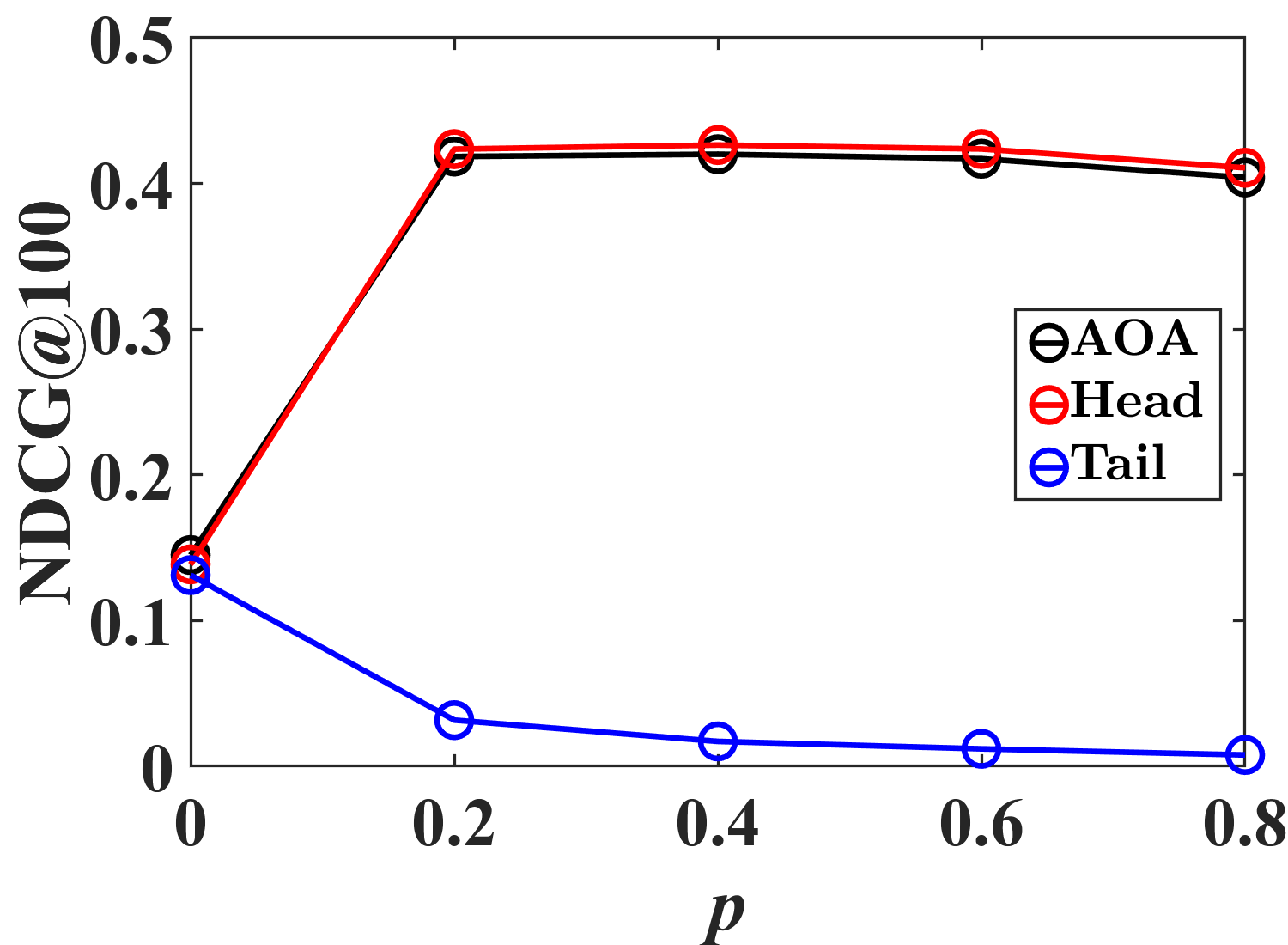} &
\includegraphics[width=0.22\textwidth]{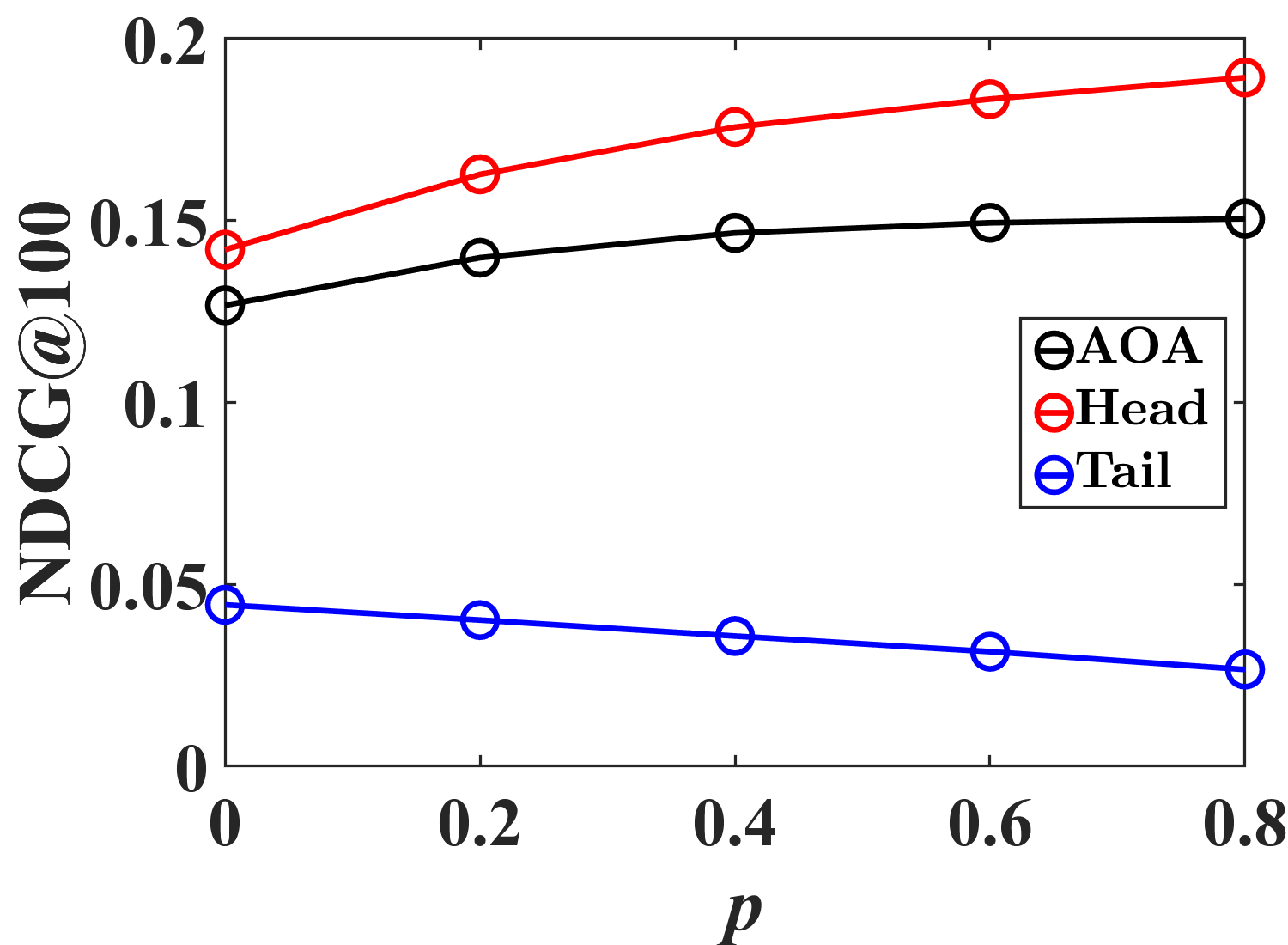} \\
(a) ML-20M & (b) Yelp2018 \\
\end{tabular}
\vspace{-2mm}
\caption{NDCG@100 of DLAE over various $\mathbf{p}$ for dropout regularization on two datasets, ML-20M and Yelp2018.}
\vspace{-2mm}
\label{fig:various_dropout}
\end{figure}

To analyze the impact of the hyperparameters, we report the results on the strong generalization protocol for two datasets, ML-20M and Yelp2018. NDCG@100 is used as the default metric. In Figures~\ref{fig:various_xi} and~\ref{fig:various_dropout}, we fix the L2 regularization coefficient $\lambda$ to 100. Note that the same trend is shown in other metrics and $\lambda$.

\noindent
\textbf{Diagonal constraints}. Figure~\ref{fig:various_xi} shows the performance of RLAE over varying $\xi$. As $\xi$ goes down, the impact of the diagonal constraints becomes stronger. In particular, $\xi=0$ indicates the zero-diagonal constraints, meaning that the diagonal constraints are applied to all items. The best performance is shown at a specific $\xi$ value, higher than when the zero-diagonal constraints are applied on both datasets. In addition, the performance of the tail items gradually improves as the $\xi$ value increases. This observation implies that the diagonal constraints suppress the collaborative signals of unpopular items. Moreover, we observe that the diagonal constraints do not affect most items at the optimal $\xi$ value. Specifically, RLAE removes 81.31\% and 78.76\% of the diagonal constraints on the ML-20M and Yelp2018 datasets, respectively.

\begin{figure}[t!]
\centering
\begin{tabular}{cc}
\includegraphics[width=0.22\textwidth]{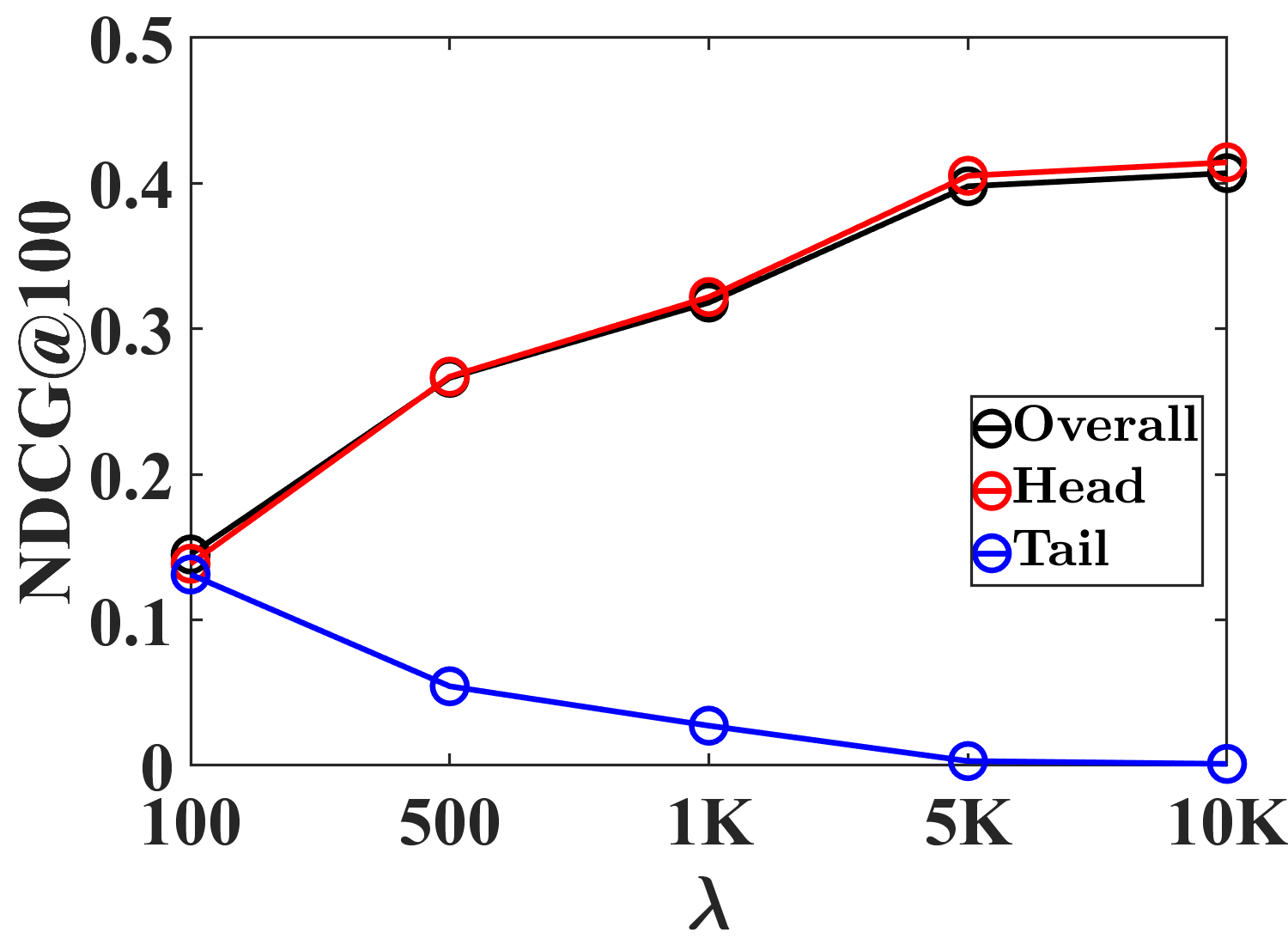} &
\includegraphics[width=0.22\textwidth]{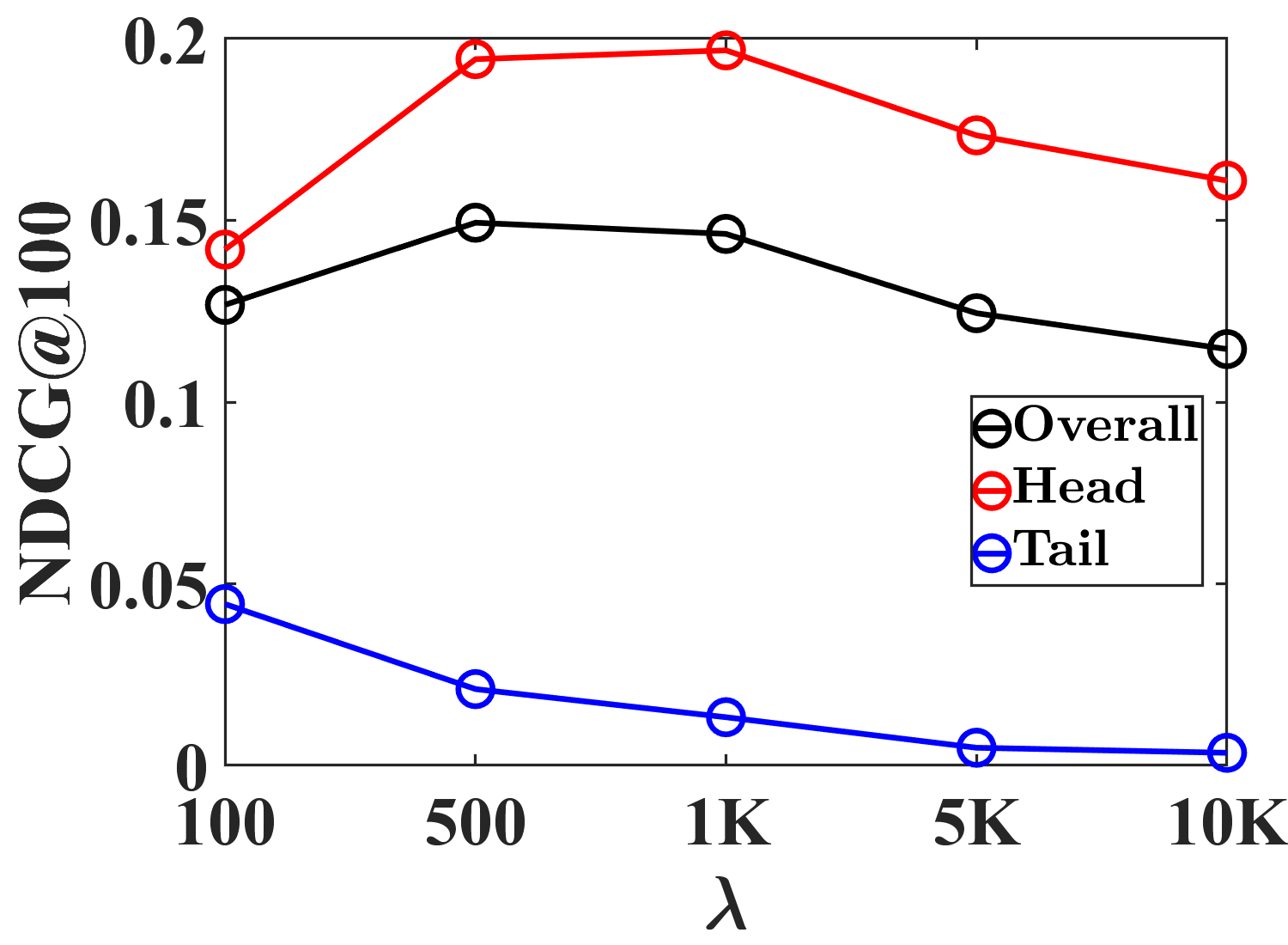} \\
(a) ML-20M & (b) Yelp2018 \\
\end{tabular}
\vspace{-2mm}
\caption{NDCG@100 of LAE over various $\boldsymbol{\lambda}$ for L2 regularization on two datasets, ML-20M and Yelp2018.}
\vspace{-2mm}
\label{fig:various_L2}
\end{figure}

\noindent
\textbf{L2 regularization}. Figure~\ref{fig:various_L2} depicts the performance of LAE over various $\lambda$. We use LAE to analyze only the impact of L2 regularization. We highlight the two observations: (1) As $\lambda$ increases, the performance of the head item group tends to increase while the performance of the tail item group tends to decrease. Therefore, it is necessary to adjust the modest L2 regularization to consider the performance balance of both item groups. (2) $\lambda=10K$ shows the best performance on the ML-20M dataset, but the best performance is shown at $\lambda=500$ on the Yelp2018 dataset. Due to the relatively low popularity bias, low-ranked PCs contain more meaningful information on Yelp2018, resulting in the best performance at relatively low $\lambda$. In brief, the optimal value of $\lambda$ depends on the popularity bias.

\noindent
\textbf{Dropout regularization}. Figure~\ref{fig:various_dropout} shows the performance of DLAE over varying dropout ratio $p$. We observe similar trends for dropout and diagonal constraints. As the $p$ increases, the performance of head items tends to increase while the performance of tail items consistently decreases for both datasets. Moreover, we observe a correlation between the popularity bias of the dataset and the value of optimal $p$. The larger the item popularity bias, the stronger the impact of dropout regularization. Thus, the variation of performance with $p$ on the ML-20M dataset is more sensitive, with optimal performance occurring at a relatively low $p$ compared to the Yelp2018 dataset.

\section{Related Work}~\label{sec:relatedwork}
We briefly review existing CF models into two groups: linear and non-linear. We also discuss theoretical analyses for understanding linear models.

\vspace{1mm}
\noindent
\textbf{Linear models}. They are categorized into two groups, \emph{latent factor models} and \emph{neighborhood-based models}. Latent factor models~\cite{HuKV08WMF1, PanZCLLSY08WMF2, ZhouWSP08ALSWR, Koren08SVD++} factorize an entire matrix into a combination of user and item vectors. Pioneering by~\cite{NingK11SLIM}, linear autoencoder models are formulated by the regression model for the item neighborhood-based approach~\cite{SarwarKKR01CFItemKNN}. Recent studies~\cite{Steck19EASE, SteckDRJ20ADMMSLIM, JeunenBG20CEASE, SteckL21Higher} calculate an item-item weight matrix via convex optimization with L2 regularization and zero-diagonal constraints. Furthermore, EDLAE~\cite{Steck20edlae} utilizes advanced regularization derived from random dropout.

\vspace{1mm}
\noindent
\textbf{Non-linear models}. With the blossom of deep learning, non-linear models have been widely used for recommendation. Like linear models, non-linear models are broadly categorized into latent factor models and autoencoder models. Non-linear latent factor models include MF-based neural models~\cite{RendleFGS09BPRMF, HeLZNHC17NeuMF, RaoYRD15GRMF} and GCN-based models~\cite{0001DWLZ020LightGCN, ChenWHZW20LRGCCF, Wang0WFC19NGCF, 0002JP21LTOCF, ShenWZSZLL21GFCF}. In addition, non-linear autoencoder models~\cite{WuDZE16CDAE, LiangKHJ18MultVAE, ShenbinATMN20RecVAE, LobelLGC20RaCT} utilize a bottleneck architecture consisting of an encoder and a decoder, where the hidden layer represents a non-linear activation function.

\vspace{1mm}
\noindent
\textbf{Theoretical analysis for linear models}. Several studies~\cite{XuRKKA21WSDM, ShenWZSZLL21GFCF, JinLGLCZ21KDD} have recently conducted in-depth analyses of linear models. \cite{XuRKKA21WSDM} performed a theoretical analysis on product embedding using skip-gram negative sampling, and \cite{ShenWZSZLL21GFCF} considered graph signal processing on the GCN-based methodology for the recommendation. In addition, \cite{JinLGLCZ21KDD} compares low-rank linear autoencoders with Tikhonov regularization and a closed-form solution of MF. To the best of our knowledge, no existing study explores diagonal constraints in linear autoencoders.

\section{Conclusion}\label{sec:conclusion}
This paper provided a theoretical understanding of linear autoencoder models. Their solutions can be decomposed into two components: an objective function with regularization and zero-diagonal constraints. To the best of our knowledge, no existing study thoroughly investigates the relationship between item popularity and two components of linear autoencoders. Regularization emphasizes high-ranked PCs, affecting strong collaborative signals. Meanwhile, the diagonal constraints weaken the impact of low-ranked PCs. Because low-ranked PCs are highly related to the collaborative signals for unpopular items, the zero-diagonal constraints are not always helpful for enhancing performances, especially for long-tail items. Motivated by our analyses, we suggested linear autoencoder models by relaxing diagonal constraints. Experimental results extensively demonstrated that our models are comparable to or better than existing linear autoencoder and non-linear models using two evaluation protocols on six benchmark datasets.

\begin{acks}
This work was supported by Institute of Information \& communications Technology Planning \& Evaluation (IITP) grant funded by the Korea government (MSIT) (No. 2022-0-00680, 2022-0-01045, 2019-0-00421, 2021-0-02068, and IITP-2023-2020-0-01821).
\end{acks}

\bibliographystyle{ACM-Reference-Format}
\balance
\bibliography{references}

\end{document}